\pdfoutput=1
\documentclass[12pt,a4paper]{article}

\usepackage{ifthen} 
\newboolean{pdflatex}
\setboolean{pdflatex}{true} 

\newboolean{articletitles}
\setboolean{articletitles}{true} 

\newboolean{uprightparticles}
\setboolean{uprightparticles}{false} 

\def\paperauthors{LHCb collaboration} 
\def\paperasciititle{Angular analysis of B0 -> D*- Ds*+ with Ds*+ -> Ds+ gamma decays} 
\def\papertitle{Angular analysis of $B^0 \to D^{*-}D_s^{*+}$ with $D_s^{*+} \to D_s^+ \gamma$ decays} 
\def\paperkeywords{{High Energy Physics}, {LHCb}} 
\def\papercopyright{\the\year\ CERN for the benefit of the LHCb collaboration} 
\def\paperlicence{CC BY 4.0 licence}
\def\paperlicenceurl{https://creativecommons.org/licenses/by/4.0/}

\def\BdDstDsst{\mbox{\ensuremath{B^0 \to D^{*-} D_s^{*+}}}\xspace}
\def\BdDstTauNu{\mbox{\ensuremath{B^0 \to D^{*-} \tau^+ \nu_\tau}}\xspace}
\def\TauThreeProng{\mbox{\ensuremath{\tau^+ \to \pi^+ \pi^+ \pi^- \bar{\nu}_\tau}}\xspace}
\def\BdDstDs{\mbox{\ensuremath{B^0 \to D^{*-} D_s^+}}\xspace}

\def\mDstDs{\mbox{\ensuremath{m(D^{*-}D_s^+)}}\xspace}

\def\fL{\ensuremath{f_{\rm L}}\xspace}

\def\aHz{\ensuremath{|H_0|}\xspace}
\def\aHp{\ensuremath{|H_+|}\xspace}
\def\aHm{\ensuremath{|H_-|}\xspace}

\def\pHp{\ensuremath{\phi_+}\xspace}
\def\pHm{\ensuremath{\phi_-}\xspace}

\def\fLsyst{0.578 \pm 0.010 \pm 0.011} 
\def\ndstdsstat{20890 \pm 178} 
\def\ndstdsststat{37415 \pm 361} 
\def\nbstodstdsstat{261 \pm 30} 
\def\BFratio{2.045 \pm 0.022 \pm 0.071} 
\def\BsBFratio{0.049 \pm 0.006 \pm 0.003 \pm 0.002} 
\def\effratio{1.142 \pm 0.034} 
 
\def\ndstdsstfullreco{6457 \pm 116} 
\def\Hzamp{0.760 \pm 0.007 \pm 0.007}

\def\Hmamp{0.195 \pm 0.022 \pm 0.032} 
\def\Hmphi{0.108 \pm 0.170 \pm 0.051} 
\def\Hpphi{-0.046 \pm 0.102 \pm 0.020} 
\def\Hpamp{0.620 \pm 0.011 \pm 0.013}

\usepackage[top=1in, bottom=1.25in, left=1in, right=1in]{geometry}

\usepackage{microtype}
\usepackage{lineno}  
\usepackage{xspace} 
\usepackage{caption} 

\usepackage{graphicx}  
\usepackage{color}
\usepackage{colortbl}
\graphicspath{{./figs/}} 

\usepackage{amsmath} 
\usepackage{amssymb}
\usepackage{amsfonts}
\usepackage{upgreek} 

\newcommand*\patchAmsMathEnvironmentForLineno[1]{%
\expandafter\let\csname old#1\expandafter\endcsname\csname #1\endcsname
\expandafter\let\csname oldend#1\expandafter\endcsname\csname
end#1\endcsname
 \renewenvironment{#1}%
   {\linenomath\csname old#1\endcsname}%
   {\csname oldend#1\endcsname\endlinenomath}%
}
\newcommand*\patchBothAmsMathEnvironmentsForLineno[1]{%
  \patchAmsMathEnvironmentForLineno{#1}%
  \patchAmsMathEnvironmentForLineno{#1*}%
}
\AtBeginDocument{%
\patchBothAmsMathEnvironmentsForLineno{equation}%
\patchBothAmsMathEnvironmentsForLineno{align}%
\patchBothAmsMathEnvironmentsForLineno{flalign}%
\patchBothAmsMathEnvironmentsForLineno{alignat}%
\patchBothAmsMathEnvironmentsForLineno{gather}%
\patchBothAmsMathEnvironmentsForLineno{multline}%
\patchBothAmsMathEnvironmentsForLineno{eqnarray}%
}

\usepackage{hyperxmp}

\usepackage[pdftex,
            pdfauthor={\paperauthors},
            pdftitle={\paperasciititle},
            pdfkeywords={\paperkeywords},
            pdfcopyright={Copyright (C) \papercopyright},
            pdflicenseurl={\paperlicenceurl}]{hyperref}

\usepackage[colorinlistoftodos,textsize=scriptsize]{todonotes}

\usepackage[bottom,flushmargin,hang,multiple]{footmisc}

\usepackage[all]{hypcap} 

\usepackage{xspace} 
\usepackage{upgreek}

\def\lhcb   {\mbox{LHCb}\xspace}

\def\MagUp {\mbox{\em Mag\kern -0.05em Up}\xspace}

\ifthenelse{\boolean{uprightparticles}}%
{
 
 \def\Pgamma      {\ensuremath{\upgamma}\xspace}

 \def\Ppi         {\ensuremath{\uppi}\xspace}

 \def\PDelta      {\ensuremath{\Delta}\xspace}                 
 \def\PXi         {\ensuremath{\Xi}\xspace}                 
 \def\PLambda     {\ensuremath{\Lambda}\xspace}                 
 \def\PSigma      {\ensuremath{\Sigma}\xspace}                 
 \def\POmega      {\ensuremath{\Omega}\xspace}                 
 \def\PUpsilon    {\ensuremath{\Upsilon}\xspace}

 \def\PB      {\ensuremath{\mathrm{B}}\xspace}                 
                  
 \def\PD      {\ensuremath{\mathrm{D}}\xspace}

 \def\PK      {\ensuremath{\mathrm{K}}\xspace}

 \def\Pb      {\ensuremath{\mathrm{b}}\xspace}                 
 \def\Pc      {\ensuremath{\mathrm{c}}\xspace}

 \def\Pi      {\ensuremath{\mathrm{i}}\xspace}

 \def\Ps      {\ensuremath{\mathrm{s}}\xspace}

 \def\thebaroffset{0.0em}
}
{
 
 \def\Pgamma      {\ensuremath{\gamma}\xspace}

 \def\Ppi         {\ensuremath{\pi}\xspace}

 \mathchardef\PDelta="7101
 \mathchardef\PXi="7104
 \mathchardef\PLambda="7103
 \mathchardef\PSigma="7106
 \mathchardef\POmega="710A
 \mathchardef\PUpsilon="7107
                  
 \def\PB      {\ensuremath{B}\xspace}                 
                  
 \def\PD      {\ensuremath{D}\xspace}

 \def\PK      {\ensuremath{K}\xspace}

 \def\Pb      {\ensuremath{b}\xspace}                 
 \def\Pc      {\ensuremath{c}\xspace}

 \def\Pi      {\ensuremath{i}\xspace}

 \def\Ps      {\ensuremath{s}\xspace}

 \def\thebaroffset{0.18em}
}
\newcommand{\offsetoverline}[2][\thebaroffset]{\kern #1\overline{\kern -#1 #2}}%

\makeatletter
\ifcase \@ptsize \relax
  \newcommand{\miniscule}{\@setfontsize\miniscule{4}{5}}
\or
  \newcommand{\miniscule}{\@setfontsize\miniscule{5}{6}}
\or
  \newcommand{\miniscule}{\@setfontsize\miniscule{5}{6}}
\fi
\makeatother

\DeclareRobustCommand{\optbar}[1]{\shortstack{{\miniscule (\rule[.5ex]{1.25em}{.18mm})}
  \\ [-.7ex] $#1$}}

\def\g      {{\ensuremath{\Pgamma}}\xspace}

\def\squark    {{\ensuremath{\Ps}}\xspace}

\def\cquark    {{\ensuremath{\Pc}}\xspace}

\def\bquark    {{\ensuremath{\Pb}}\xspace}

\def\pion   {{\ensuremath{\Ppi}}\xspace}
\def\piz    {{\ensuremath{\pion^0}}\xspace}
\def\pip    {{\ensuremath{\pion^+}}\xspace}
\def\pim    {{\ensuremath{\pion^-}}\xspace}

\def\kaon    {{\ensuremath{\PK}}\xspace}

\def\KorKbar {\kern \thebaroffset\optbar{\kern -\thebaroffset \PK}{}\xspace}

\def\Kp      {{\ensuremath{\kaon^+}}\xspace}
\def\Km      {{\ensuremath{\kaon^-}}\xspace}

\def\Dbar    {{\ensuremath{\offsetoverline{\PD}}}\xspace}
\def\D       {{\ensuremath{\PD}}\xspace}

\def\DorDbar {\kern \thebaroffset\optbar{\kern -\thebaroffset \PD}\xspace}

\def\Dzb     {{\ensuremath{\Dbar{}^0}}\xspace}
\def\Dp      {{\ensuremath{\D^+}}\xspace}
\def\Dm      {{\ensuremath{\D^-}}\xspace}

\def\DpDm    {\ensuremath{\Dp {\kern -0.16em \Dm}}\xspace}
\def\Dstar   {{\ensuremath{\D^*}}\xspace}

\def\Dstarm  {{\ensuremath{\D^{*-}}}\xspace}

\def\Dsp     {{\ensuremath{\D^+_\squark}}\xspace}

\def\Dssp    {{\ensuremath{\D^{*+}_\squark}}\xspace}

\def\B       {{\ensuremath{\PB}}\xspace}
\def\Bbar    {{\ensuremath{\offsetoverline{\PB}}}\xspace}

\def\BorBbar {\kern \thebaroffset\optbar{\kern -\thebaroffset \PB}\xspace}
\def\Bz      {{\ensuremath{\B^0}}\xspace}
\def\Bzb     {{\ensuremath{\Bbar{}^0}}\xspace}
\def\Bd      {{\ensuremath{\B^0}}\xspace}

\def\BdorBdbar {\kern \thebaroffset\optbar{\kern -\thebaroffset \Bd}\xspace}

\def\Bs      {{\ensuremath{\B^0_\squark}}\xspace}

\def\BsorBsbar {\kern \thebaroffset\optbar{\kern -\thebaroffset \Bs}\xspace}

\def\Y#1S{\ensuremath{\PUpsilon{(#1S)}}\xspace}

\def\LorLbar     {\kern \thebaroffset\optbar{\kern -\thebaroffset \PLambda}\xspace}

\def\to                 {\ensuremath{\rightarrow}\xspace}

\def\CP                {{\ensuremath{C\!P}}\xspace}

\def\AT#1     {\ensuremath{A_{\mathrm{T}}^{#1}}\xspace}           

\def\C#1      {\ensuremath{\mathcal{C}_{#1}}\xspace}                       
\def\Cp#1     {\ensuremath{\mathcal{C}_{#1}^{'}}\xspace}                    
\def\Ceff#1   {\ensuremath{\mathcal{C}_{#1}^{\mathrm{(eff)}}}\xspace}        
\def\Cpeff#1  {\ensuremath{\mathcal{C}_{#1}^{'\mathrm{(eff)}}}\xspace}       
\def\Ope#1    {\ensuremath{\mathcal{O}_{#1}}\xspace}                       
\def\Opep#1   {\ensuremath{\mathcal{O}_{#1}^{'}}\xspace}                    

\newcommand{\nospaceunit}[1]{\ensuremath{\text{#1}}}       
\newcommand{\aunit}[1]{\ensuremath{\text{\,#1}}}       

\newcommand{\tev}{\aunit{Te\kern -0.1em V}\xspace}
\newcommand{\gev}{\aunit{Ge\kern -0.1em V}\xspace}
\newcommand{\mev}{\aunit{Me\kern -0.1em V}\xspace}
\newcommand{\kev}{\aunit{ke\kern -0.1em V}\xspace}
\newcommand{\ev}{\aunit{e\kern -0.1em V}\xspace}
 
\newcommand{\mevc}{\ensuremath{\aunit{Me\kern -0.1em V\!/}c}\xspace}
\newcommand{\gevc}{\ensuremath{\aunit{Ge\kern -0.1em V\!/}c}\xspace}
\newcommand{\mevcc}{\ensuremath{\aunit{Me\kern -0.1em V\!/}c^2}\xspace}
\newcommand{\gevcc}{\ensuremath{\aunit{Ge\kern -0.1em V\!/}c^2}\xspace}

\def\mum  {\ensuremath{\,\upmu\nospaceunit{m}}\xspace}

\def\fb   {\ensuremath{\aunit{fb}}\xspace}
\def\invfb   {\ensuremath{\fb^{-1}}\xspace}

\def\ps   {\ensuremath{\aunit{ps}}\xspace}

\def\gsim{{~\raise.15em\hbox{$>$}\kern-.85em
          \lower.35em\hbox{$\sim$}~}\xspace}
\def\lsim{{~\raise.15em\hbox{$<$}\kern-.85em
          \lower.35em\hbox{$\sim$}~}\xspace}

\def\sPlot{\mbox{\em sPlot}\xspace}

\def\pt         {\ensuremath{p_{\mathrm{T}}}\xspace}

\def\ptot       {\ensuremath{p}\xspace}

\def\evtgen     {\mbox{\textsc{EvtGen}}\xspace}

\def\geant      {\mbox{\textsc{Geant4}}\xspace}

\def\photos     {\mbox{\textsc{Photos}}\xspace}

\def\pythia     {\mbox{\textsc{Pythia}}\xspace}

\def\tell1  {TELL1\xspace}
\def\ukl1   {UKL1\xspace}

\usepackage{cite} 
\usepackage{mciteplus}

\usepackage{longtable} 
\usepackage{placeins}
\usepackage{booktabs}

\begin{document}

\renewcommand{\thefootnote}{\fnsymbol{footnote}}
\setcounter{footnote}{1}


\begin{titlepage}
\pagenumbering{roman}

\vspace*{-1.5cm}
\centerline{\large EUROPEAN ORGANIZATION FOR NUCLEAR RESEARCH (CERN)}
\vspace*{1.5cm}
\noindent
\begin{tabular*}{\linewidth}{lc@{\extracolsep{\fill}}r@{\extracolsep{0pt}}}
\ifthenelse{\boolean{pdflatex}}
{\vspace*{-1.5cm}\mbox{\!\!\!\includegraphics[width=.14\textwidth]{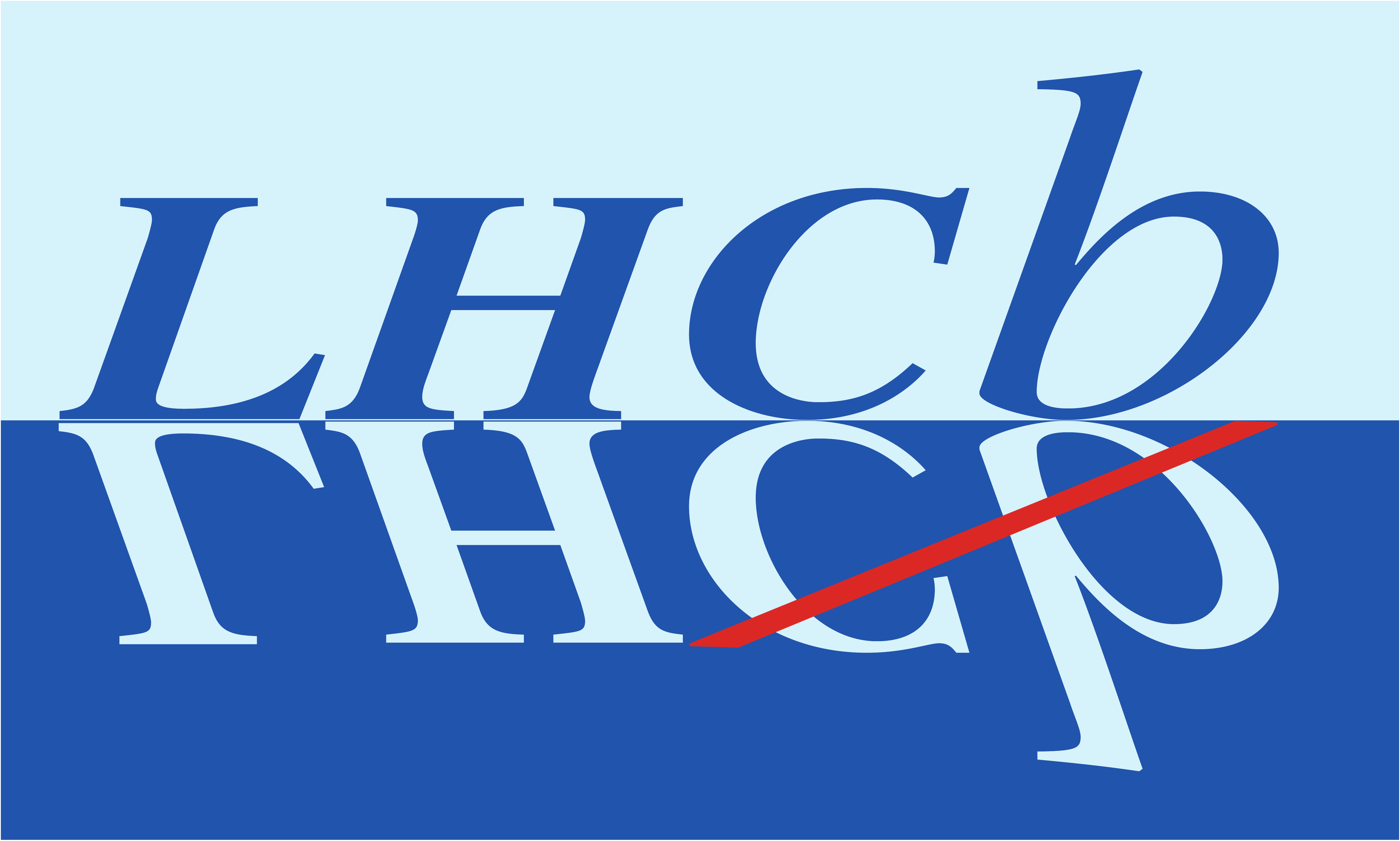}} & &}%
{\vspace*{-1.2cm}\mbox{\!\!\!\includegraphics[width=.12\textwidth]{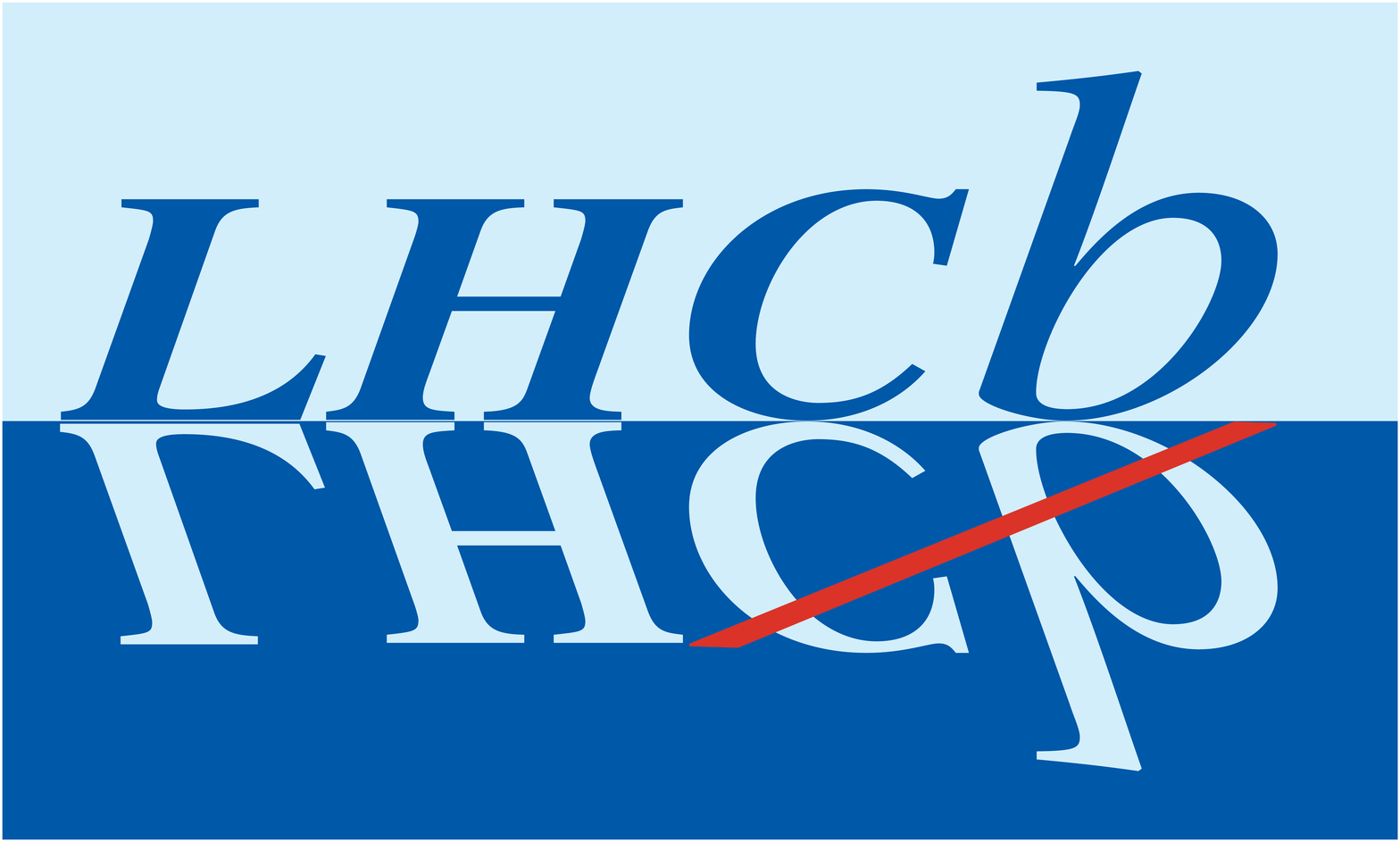}} & &}%
\\
 & & CERN-EP-2021-056 \\  
 & & LHCb-PAPER-2021-006 \\  
 & & \today \\ 
 & & \\
\end{tabular*}

\vspace*{1.0cm}

{\normalfont\bfseries\boldmath\huge
\begin{center}
  \papertitle 
\end{center}
}

\vspace*{2.0cm}

\begin{center}
\paperauthors\footnote{Authors are listed at the end of this paper.}
\end{center}

\vspace{\fill}

\begin{abstract}
  \noindent
  The first full angular analysis of the $B^0 \to D^{*-} D_s^{*+}$ decay is performed using 6 fb$^{-1}$ of $pp$ collision data collected with the LHCb experiment at a centre-of-mass energy of 13 TeV. The \mbox{$D_s^{*+} \to D_s^+ \gamma$} and $D^{*-} \to \Dzb \pi^-$ vector meson decays are used with the subsequent \mbox{$D_s^+ \to K^+ K^- \pi^+$} and $\Dzb \to K^+ \pi^-$ decays. All helicity amplitudes and phases are measured, and the longitudinal polarisation fraction is determined to be $f_{\rm L} = \fLsyst$  with world-best precision, where the first uncertainty is statistical and the second is systematic. The pattern of helicity amplitude magnitudes is found to align with expectations from quark-helicity conservation in $B$ decays. The ratio of branching fractions \mbox{$[\mathcal{B}(B^0 \to D^{*-} D_s^{*+}) \times \mathcal{B}(D_s^{*+} \to D_s^+ \gamma)]/\mathcal{B}(B^0 \to D^{*-} D_s^+)$} is measured to be \mbox{$\BFratio$} with world-best precision. In addition, the first observation of the Cabibbo-suppressed \mbox{$\Bs \to D^{*-} D_s^+$} decay is made with a significance of seven standard deviations. The branching fraction ratio \mbox{$\mathcal{B}(\Bs \to D^{*-} D_s^+)/\mathcal{B}(B^0 \to D^{*-} D_s^+)$} is measured to be \mbox{$\BsBFratio$}, where the third uncertainty is due to limited knowledge of the ratio of fragmentation fractions.
\end{abstract}

\vspace*{2.0cm}

\begin{center}
  Published in JHEP 06 (2021) 177.
\end{center}

\vspace{\fill}

{\footnotesize 
\centerline{\copyright~\papercopyright. \href{\paperlicenceurl}{\paperlicence}.}}
\vspace*{2mm}

\end{titlepage}


\newpage
\setcounter{page}{2}
\mbox{~}
%
%
%
%


\renewcommand{\thefootnote}{\arabic{footnote}}
\setcounter{footnote}{0}

\cleardoublepage


\pagestyle{plain} 
\setcounter{page}{1}
\pagenumbering{arabic}


\section{Introduction}
\label{sec:Introduction}

The \BdDstDsst decay involves the production of two vector charm mesons from a pseudoscalar \Bz parent. This process exhibits a polarisation structure, where three complex helicity amplitudes $H_0$, $H_+$, and $H_-$ contribute to the total decay rate. These amplitudes correspond to the relative orientation of the linear polarisation vectors of the two vector mesons. Parity-even ($\parallel$) and parity-odd ($\perp$) transversity amplitudes can also be defined in terms of $H_+$ and $H_-$, namely $A_{\parallel,\perp} = (H_+ \pm H_-)/\sqrt{2}$. The helicity amplitudes can interfere, with interference governed by the strong phases of the transverse components, \pHp and \pHm, relative to the phase of the longitudinal component, $\phi_0$, which is conventionally taken to be equal to zero. Therefore, five parameters in total determine the decay rate:
\begin{itemize}
    \item \aHz, the magnitude of the longitudinal amplitude;
    \item \aHp and \aHm, the magnitudes of the two transverse amplitudes;
    \item \pHp and \pHm, the phases of the transverse amplitudes relative to $H_0$.
\end{itemize}
In order to normalise the total decay rate, \mbox{$|H_0|^2 + |H_+|^2 + |H_-|^2 = f_{\rm L} + f_{\rm T} = 1$} is required, where $f_{\rm L} \equiv |H_0|^2$ is the longitudinal polarisation fraction and $f_{\rm T} \equiv |H_+|^2 + |H_-|^2$ is the transverse polarisation fraction. The current world average for \fL is $0.52 \pm 0.05$~\cite{Aubert:2003jj,PDG2020}, while theoretical predictions cover a similar range~\cite{Kramer_Palmer,Ahmed:2000ad, PhysRevLett.76.3898,PhysRevD.64.094001}; the transverse helicity amplitudes have not been measured previously. The normalisation condition reduces the total number of independent observables to four, where the additional observable is absorbed into the absolute branching fraction of the decay which is not measured. Measuring the relative magnitudes of the helicity amplitudes offers a test of quark-helicity conservation in this tree-level decay involving a $b \to c$ quark transition. In such decays, a $|H_0| > |H_+| > |H_-|$ hierarchy is expected~\cite{Suzuki:2001za}, where the $V-A$ nature of the weak interaction causes the longitudinal component to dominate.

The \BdDstDsst decay has a large branching fraction, \mbox{$\mathcal{B}(B^0 \to D^{*-} D_s^{*+}) = (1.77 \pm 0.14)\%$}~\cite{PDG2020}, and is thus a prominent background in \BdDstTauNu analyses that exploit the hadronic three-prong \TauThreeProng mode in order to measure the ratio $R(D^*) \equiv \mathcal{B}(B^0 \to D^{*-} \tau^+ \nu_\tau)/\mathcal{B}(B^0 \to D^{*-}\ell^{+}\nu_\ell)$~\cite{LHCb-PAPER-2017-027} or the angular coefficients of the \BdDstTauNu decay~\cite{Hill:2019zja}. 
Such a background arises when the neutral particle produced in the \Dssp decay is not reconstructed, and the \Dsp meson decays to three pions plus additional non-reconstructed particles.

Using data corresponding to an integrated luminosity of 6\invfb collected at a centre-of-mass energy of 13 TeV with the LHCb experiment between 2015 and 2018, \BdDstDsst with $D_s^{*+} \to D_s^+ \gamma$ decays are reconstructed via the \mbox{$D^{*-} \to (\Dzb \to \Kp \pim) \pim$} and \mbox{$D_s^+ \to \Kp \Km \pip$} channels; the inclusion of charge-conjugate processes is implied throughout. Partially reconstructed decays, where the photon is not considered in the invariant-mass calculation, are used in a fit to the \mDstDs distribution to measure \fL. Fully reconstructed decays are then considered in a subsequent angular analysis to measure the remaining helicity observables. Measurements are performed under the assumption that both the $\Dzb \pim$ and $D_s^+ \gamma$ systems are pure vector, as no evidence for a scalar contribution is found in the $m(\Dzb \pim)$ distribution in data and no scalar component is permitted in $m(\Dsp \g)$ due to the photon angular momentum. The analysis includes an improved measurement of \fL and first measurements of the transverse helicity amplitude magnitudes and phases. 

The data sample is also used to measure the ratio of branching fractions \mbox{$\mathcal{R} \equiv [\mathcal{B}(B^0 \to D^{*-} D_s^{*+}) \times \mathcal{B}(D_s^{*+} \to D_s^+ \gamma)]/\mathcal{B}(B^0 \to D^{*-} D_s^+)$}, where the current value of $\mathcal{R} = 2.07 \pm 0.33$ is calculated using world-average branching fractions taken from Ref.~\cite{PDG2020}. In addition, a measurement of the previously unobserved Cabibbo-suppressed \mbox{$\Bs \to D^{*-} D_s^+$} decay is performed and the ratio of branching fractions \mbox{$\mathcal{B}(\Bs \to D^{*-} D_s^+)/\mathcal{B}(B^0 \to D^{*-} D_s^+)$} determined.

The formalism adopted is described in Sect.~\ref{sec:formalism}, essential details of the LHCb detector and simulation are given in Sect.~\ref{sec:Detector}, and the event selection is outlined in Sect.~\ref{sec:Selection}. The longitudinal polarisation fraction and ratios of branching fractions are measured in Sect.~\ref{sec:fL_Fit}, and the remaining helicity observables are measured in Sects.~\ref{sec:sFit}--\ref{sec:Fit}. Systematic uncertainties are determined in Sect.~\ref{sec:Syst}, and final results and conclusions are presented in Sect.~\ref{sec:Results}. 

\section{Angular decay rate formalism}
\label{sec:formalism}

\begin{figure}[!h]
  \begin{center}
   \includegraphics[width=0.99\linewidth]{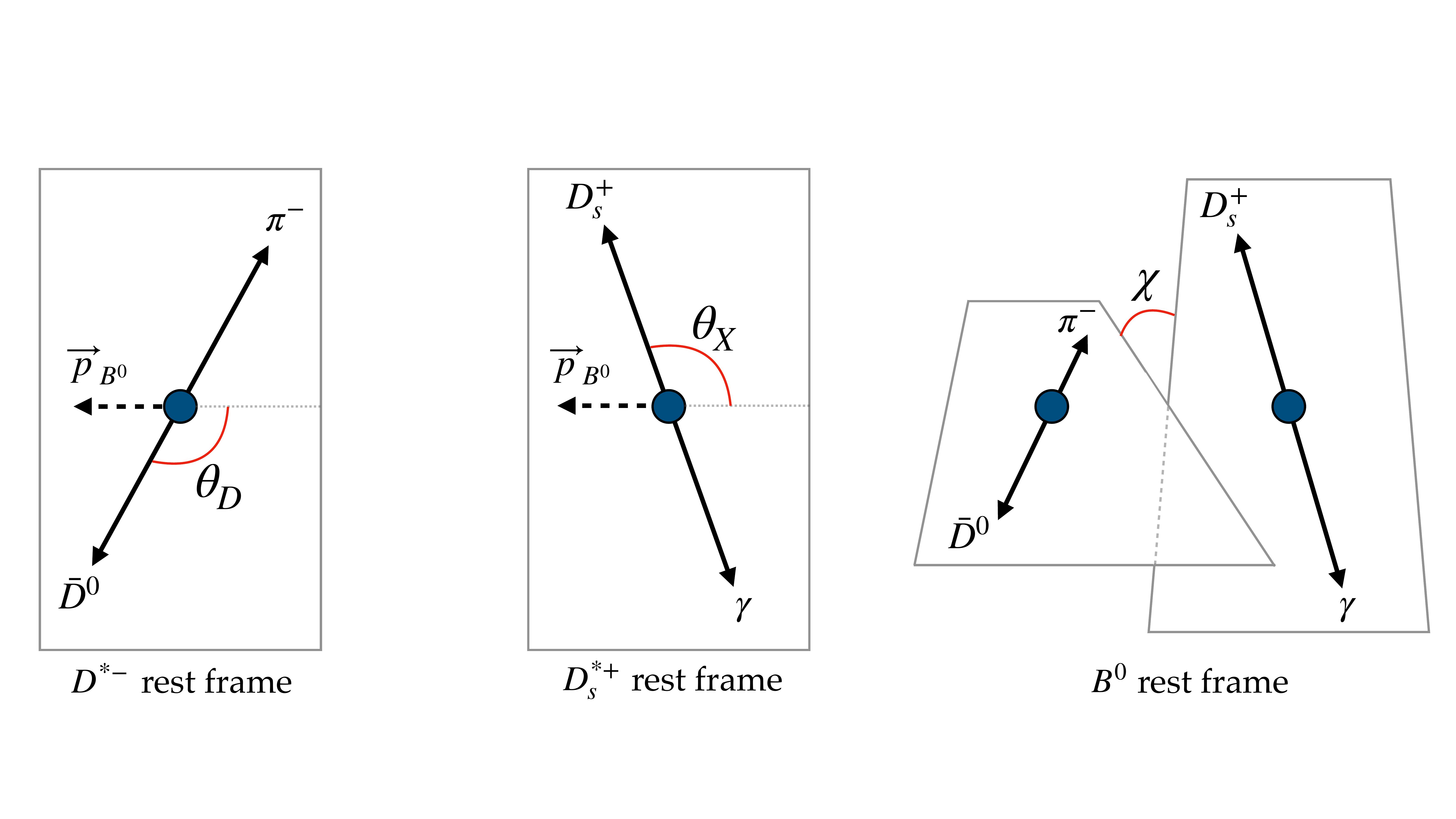}
   \end{center}
   \caption{Illustration of the \BdDstDsst decay angles.
  \label{fig:decay_angles}}
\end{figure}

The \BdDstDsst decay rate is a function of three decay angles, $\theta_D$, $\theta_X$, and $\chi$, where $\theta_D$ is the angle between the \Dzb meson and the direction opposite the \Bz momentum vector in the \Dstarm rest frame, $\theta_X$ is the angle between the \Dsp meson and the direction opposite the \Bz momentum vector in the \Dssp rest frame, and $\chi$ is the angle between the two decay planes as defined in the \Bz rest frame. The angles are illustrated in Fig.~\ref{fig:decay_angles}, and are explicitly defined as follows
\begin{align}
\cos\theta_D &= \Big(\hat{p}_{D^0}^{(D^{*-})} \Big) \cdot \Big(\hat{p}_{D^{*-}}^{(B^0)} \Big) = \Big(\hat{p}_{D^0}^{(D^{*-})} \Big) \cdot \Big(-\hat{p}_{B^0}^{(D^{*-})} \Big)\,, \notag\\
\cos\theta_X &= \Big(\hat{p}_{D_s^+}^{(D_s^{*+})} \Big) \cdot \Big(\hat{p}_{D_s^{*+}}^{(B^0)} \Big) = \Big(\hat{p}_{D_s^+}^{(D_s^{*+})} \Big) \cdot \Big(-\hat{p}_{B^0}^{(D_s^{*+})} \Big)\,, \notag\\
\cos\chi &= \Big(\hat{p}_{D_s^+}^{(B^0)} \times \hat{p}_{\gamma}^{(B^0)} \Big) \cdot \Big(\hat{p}_{D^0}^{(B^0)} \times \hat{p}_{\pi^-}^{(B^0)} \Big)\,,\\
\sin\chi^{\Bz} &= -\Big[ \Big(\hat{p}_{D_s^+}^{(B^0)} \times \hat{p}_{\gamma}^{(B^0)} \Big) \times \Big(\hat{p}_{\Dzb}^{(B^0)} \times \hat{p}_{\pi^-}^{(B^0)} \Big) \Big] \cdot \hat{p}_{D^{*-}}^{(B^0)}\,, \notag\\
\sin\chi^{\Bzb} &= +\Big[ \Big(\hat{p}_{D_s^-}^{(\Bzb)} \times \hat{p}_{\gamma}^{(\Bzb)} \Big) \times \Big(\hat{p}_{D^0}^{(\Bzb)} \times \hat{p}_{\pi^-}^{(\Bzb)} \Big) \Big] \cdot \hat{p}_{D^{*+}}^{(\Bzb)}\,, \notag
\end{align}
where the $\hat{p}_{X}^{(Y)}$ are unit vectors describing the direction of a particle $X$ in the rest frame of the system $Y$. In the $B^0$ rest frame, the angular definition for the \Bzb decay is a charge-parity (\CP) transformation of that for the \Bz decay. The sign of $\sin\chi$ is negative for \Bz candidates and positive for $\Bzb$ candidates, where the $B$-meson flavour is tagged by the \Dstar-meson charge. This formalism is the same as that adopted in other LHCb angular analyses such as that of $B \to K^* \mu^+ \mu^-$ decays~\cite{LHCb-PAPER-2015-051,LHCb-PAPER-2020-041}.

The full three-dimensional differential decay rate expressed in terms of the helicity amplitudes is given by~\cite{Kramer_Palmer}
\begin{align}
&\frac{d^{3} \Gamma}{d \cos \theta_{D} d \cos \theta_{X} d \chi} \propto \frac{9}{8}\bigg\{\cos ^{2} \theta_{D} \sin ^{2} \theta_{X}\left|H_{0}\right|^{2}+\frac{1}{4} \sin ^{2} \theta_{D}\left(1+\cos ^{2} \theta_{X}\right)\left(\left|H_{+}\right|^{2}+\left|H_{-}\right|^{2}\right) \notag \\ \notag \\
&-\frac{1}{2} \sin ^{2} \theta_{D} \sin ^{2} \theta_{X}\left[\cos 2 \chi \operatorname{Re}\left(H_{+} H_{-}^{*}\right)-\sin 2 \chi \operatorname{Im}\left(H_{+} H_{-}^{*}\right)\right] \label{eq:decay_rate} \\ \notag \\ 
&-\frac{1}{4} \sin 2 \theta_{D} \sin 2 \theta_{X}\left[\cos \chi \operatorname{Re}\left(H_{+} H_{0}^{*}+H_{-} H_{0}^{*}\right)-\sin \chi \operatorname{Im}\left(H_{+} H_{0}^{*}-H_{-} H_{0}^{*}\right)\right] \bigg\} \notag\,.
\end{align}

\section{LHCb detector and simulation}
\label{sec:Detector}

The \lhcb detector~\cite{LHCb-DP-2008-001,LHCb-DP-2014-002} is a single-arm forward
spectrometer covering the \mbox{pseudorapidity} range $2<\eta <5$,
designed for the study of particles containing \bquark- or \cquark-quarks. The detector includes a high-precision tracking system
consisting of a silicon-strip vertex detector surrounding the $pp$
interaction region, a large-area silicon-strip detector located
upstream of a dipole magnet with a bending power of about
$4{\mathrm{\,Tm}}$, and three stations of silicon-strip detectors and straw
drift tubes placed downstream of the magnet.
The tracking system provides a measurement of the momentum, \ptot, of charged particles with
a relative uncertainty that varies from 0.5\% at low momentum to 1.0\% at 200\gevc.
The minimum distance of a track to a primary $pp$ collision vertex (PV), the impact parameter (IP), 
is measured with a resolution of $(15+29/\pt)\mum$,
where \pt is the component of the momentum transverse to the beam, in\,\gevc.
Different types of charged hadrons are distinguished using information
from two ring-imaging Cherenkov detectors. Photons, electrons and hadrons are identified by a calorimeter system consisting of
scintillating-pad and preshower detectors, an electromagnetic and a hadronic calorimeter. Muons are identified by a
system composed of alternating layers of iron and multiwire
proportional chambers.

The online event selection is performed by a trigger, 
which consists of a hardware stage, based on information from the calorimeter and muon
systems, followed by a software stage, which applies a full event
reconstruction. At the hardware trigger stage, events are required to have a muon with high \pt or a
  hadron, photon or electron with high transverse energy in the calorimeters. For hadrons,
  the transverse energy threshold is 3.5\gev.
  The software trigger requires a two-, three- or four-track
  secondary vertex with a significant displacement from any primary
  $pp$ interaction vertex. At least one charged particle
  must have a transverse momentum $\pt > 1.6\gevc$ and be
  inconsistent with originating from any PV.
  A multivariate algorithm is used for
  the identification of secondary vertices consistent with the decay
  of a \bquark hadron. In the offline selection, trigger information is associated with reconstructed particles. Selection requirements can therefore be made on the trigger selection itself and on whether the decision was due to the signal candidate, other particles produced in the $pp$ collision, or an overlap of both.

Simulation is required to model the effects of the detector acceptance and the
  imposed selection requirements.
  In the simulation, $pp$ collisions are generated using
  \pythia~\cite{Sjostrand:2007gs,*Sjostrand:2006za} with a specific \lhcb configuration~\cite{LHCb-PROC-2010-056}.
  Decays of unstable particles
  are described by \evtgen~\cite{Lange:2001uf}, in which final-state
  radiation is generated using \photos~\cite{davidson2015photos}.
  The interaction of the generated particles with the detector, and its response,
  are implemented using the \geant
  toolkit~\cite{Allison:2006ve, *Agostinelli:2002hh} as described in
  Ref.~\cite{LHCb-PROC-2011-006}. The underlying $pp$ interaction is reused multiple times, with an independently generated signal decay for each~\cite{LHCb-DP-2018-004}. In addition, the \mDstDs distributions of pure longitudinal and transverse polarised \BdDstDsst decays are studied using fast-simulated samples generated with the \mbox{\textsc{RapidSim}} package~\cite{Cowan:2016tnm}, where an LHCb momentum resolution configuration is used to smear the generated four-momenta. The same tool is used to study the \mDstDs distributions of various background contributions from decays involving higher-excited charm mesons.

\section{Event selection}
\label{sec:Selection}

Candidate \BdDstDs decays are reconstructed through the $D^{*-} \to (\Dzb \to \Kp \pim) \pim$ and $D_s^+ \to \Kp \Km \pip$ channels. The tracks of the final-state particles are required to have a good quality, fulfil loose particle identification (PID) criteria, and have a high $\chi^2_{\text{IP}}$ value with respect to any PV, where $\chi^2_{\text{IP}}$ is defined as the difference in the vertex-fit $\chi^2$ of a given PV reconstructed with and without the particle being considered. The reconstructed masses of the \Dzb and $D_s^+$ candidates are required to lie inside mass windows of
$\pm 20$\mevcc around their known values~\cite{PDG2020}. The $D^{*-}$ candidate mass is required to be within $\pm 40$\mevcc of the known value~\cite{PDG2020}, while the difference in mass between the $D^{*-}$ and \Dzb candidates is required to be in the range 140--150\mevcc. In combination with the track PID cuts, these narrow mass windows reduce potential backgrounds from misidentified decays such as $B^0 \to D^{*-} D^+$ to negligible levels.

The $B^0$ candidate is reconstructed by combining the $D^{*-}$ and $D_s^+$ candidates to form a
common vertex.  If multiple PVs are reconstructed in the same event, the PV for which the $B^0$ candidate has the lowest $\chi^2_{\text{IP}}$ is assigned as the associated PV. The \pt of the $B^0$ candidate is required to be larger than 5\gevc, and the $\chi^2_{\text{IP}}$ of the $B^0$ candidate for the associated PV is required
to be small. To suppress combinatorial background and background from decays involving the production of a $D^{*-}$ and three prompt tracks, the flight distance of the $D_s^+$ candidate along the beam axis is required to be different from zero by more than one standard deviation, considering both the origin and decay-vertex uncertainties of the $D_s^+$ candidate. To suppress combinatorial background from combinations of tracks originating from the PV, the decay time of the $B^0$ candidate is required to be larger than 0.2\ps. To improve the invariant-mass resolution, a kinematic fit is performed to the decay chain~\cite{Hulsbergen:2005pu}, the $B^0$ candidate is constrained to originate from the PV and the $D_s^+$ and \Dzb masses are constrained to their known values. Candidates are retained if the resulting invariant mass of the $D^{*-}D_s^+$ combination falls within the 4900--5500\mevcc range, which includes the region occupied by partially reconstructed \BdDstDsst decays when the neutral particle produced in the $D_s^{*+}$ decay is not reconstructed. This sample is considered in Sect.~\ref{sec:fL_Fit}, where a fit to the $m(D^{*-}D_s^+)$ distribution of candidates is used to measure \fL.

A subsample of fully reconstructed \BdDstDsst candidates is selected by combining $D_s^+$ candidates from the above dataset with photons. The difference between the $D_s^{*+}$ and $D_s^+$ candidate masses is required to be in the range 120--180\mevcc, and the photon is required to have a \pt larger than 500\mevc. Each $D_s^{*+}$ candidate is then recombined with the corresponding $D^{*-}$ candidate from the above dataset to form a $B^0$ candidate, where candidates in the invariant-mass range 5150--5500\mevcc are retained. Fully reconstructed candidates with $m(D^{*-}D_s^+)$ values greater than 5240\mevcc are vetoed to remove $B^0 \to D^{*-} D_s^+$ decays where a random photon is combined with the $D_s^+$  candidate. This dataset is used in Sect.~\ref{sec:Fit} to measure the remaining helicity observables in an angular analysis.

\section{Measurement of \boldmath{$f_{\rm L}$} and branching fraction ratios}
\label{sec:fL_Fit}

The longitudinal polarisation fraction, $f_{\rm L}$, determines the fractional contribution of the $H_0$ helicity amplitude to the total \BdDstDsst decay rate. The longitudinal and transverse amplitudes contribute to the one-dimensional differential decay rate in $\cos\theta_X$ as follows,
\begin{align}
    \frac{d\Gamma}{d\cos\theta_X} &\propto \frac{3}{4}\bigg[ |H_0|^2 (1 - \cos^2\theta_X) + \frac{1}{2} (|H_+|^2 + |H_-|^2) (1 + \cos^2\theta_X) \bigg] \nonumber \\
    &= \frac{3}{4}\bigg[ f_{\rm L} (1 - \cos^2\theta_X) + \frac{(1 - f_{\rm L})}{2} (1 + \cos^2\theta_X) \bigg]\,, \label{eq:costheta_X_decay_rate}
\end{align}
which is obtained from Eq.~(\ref{eq:decay_rate}) via a definite integral over $\cos\theta_D$ and $\chi$. Experimentally, the integral over $\cos\theta_D$ and $\chi$ must also include the acceptance in these angles. However, the acceptance is predominantly linear for both angles, as shown in Figs.~\ref{fig:costheta_D_X_acceptance} and~\ref{fig:chi_acceptance}, such that no significant residual dependence remains after the integration.
Due to a common dependence on photon kinematics, the angle $\cos\theta_X$ and the invariant mass of the $D^{*-}D_s^+$ system are strongly negatively correlated, as illustrated in Appendix~\ref{app:m_DstDs_costheta_X} in Fig.~\ref{fig:m_DstDs_vs_costheta_X}. More positive values of $\cos\theta_X$ correspond to higher momentum photons and thus lower values of $m(D^{*-} D_s^+)$. As a result, the different $\cos\theta_X$ shapes for longitudinal and transverse polarised \BdDstDsst decays manifest in corresponding $m(D^{*-}D_s^+)$ distributions with different parabolic forms, as shown in Appendix~\ref{app:m_DstDs_costheta_X} in Fig.~\ref{fig:costheta_X_1D_plots}. This feature enables $f_{\rm L}$ to be measured using a binned maximum-likelihood fit to the $m(D^{*-}D_s^+)$ distribution in data, where the total \BdDstDsst contribution is modelled by the sum of probability density functions (PDFs) for the longitudinal and transverse components with relative fractions $f_{\rm L}$ and $1 - f_{\rm L}$. Determining $f_{\rm L}$ via an $m(D^{*-}D_s^+)$ fit enables partially reconstructed \BdDstDsst decays to be used, which increases the sample size by avoiding efficiency losses due to the limited photon reconstruction efficiency of the LHCb detector. 

Due to the presence of \BdDstDs decays in the same sample, a measurement of the branching fraction ratio
\begin{equation}
\mathcal{R} \equiv \frac{\mathcal{B}(B^0 \to D^{*-} D_s^{*+}) \times \mathcal{B}(D_s^{*+} \to D_s^+ \gamma)}{\mathcal{B}(B^0 \to D^{*-} D_s^+)}
\end{equation}
can also be made. Experimentally, this quantity is defined as
\begin{align}
\label{eq:BF_ratio_def}
\mathcal{R} &= \frac{\mathcal{N}(B^0 \to D^{*-} (D_s^{*+} \to D_s^+ \gamma))}{\mathcal{N}(B^0 \to D^{*-} D_s^+)} \times \frac{\epsilon(B^0 \to D^{*-} D_s^+)}{\epsilon(B^0 \to D^{*-} (D_s^{*+} \to D_s^+ \gamma))} \notag \\[10pt]
&= \frac{\mathcal{N}(B^0 \to D^{*-} (D_s^{*+} \to D_s^+ \gamma))}{\mathcal{N}(B^0 \to D^{*-} D_s^+)} \times \xi\,,
\end{align}
where $\mathcal{N}$ denotes the yields for each decay mode, and $\xi$ is the ratio of their total reconstruction and selection efficiencies. In the case of \BdDstDsst decays, the yields and efficiencies correspond to those of partially reconstructed signal. The efficiency ratio is determined using simulated samples of \BdDstDsst and \BdDstDs decays, and is found to be \mbox{$\xi = \effratio$}, where the uncertainty quoted accounts only for the use of finite simulated samples and potential variation in the efficiency across data-taking years. This uncertainty is considered as a source of systematic uncertainty on $\mathcal{R}$.

A contribution from Cabibbo-suppressed $\Bs \to D^{*-} D_s^+$ decays is also considered in the $m(D^{*-}D_s^+)$ fit, enabling a measurement of the branching fraction ratio
\begin{equation}
r(\Bs) \equiv \frac{\mathcal{B}(\Bs \to D^{*-} D_s^{+})}{\mathcal{B}(B^0 \to D^{*-} D_s^+)}
\end{equation}
to be made. Experimentally, $r(\Bs)$ is defined as
\begin{align}
\label{eq:Bs_BF_ratio_def}
r(\Bs) &= \frac{f_d}{f_s} \times \frac{\mathcal{N}(\Bs \to D^{*-} D_s^+)}{\mathcal{N}(B^0 \to D^{*-} D_s^+)} \times \frac{\epsilon(B^0 \to D^{*-} D_s^+)}{\epsilon(\Bs \to D^{*-} D_s^+)} \notag \\[10pt]
&= \frac{f_d}{f_s} \times \frac{\mathcal{N}(\Bs \to D^{*-} D_s^+)}{\mathcal{N}(B^0 \to D^{*-} D_s^+)} \times \xi(\Bs)\,,
\end{align}
where $\mathcal{N}$ denotes the yields for each decay mode, and \mbox{$f_s/f_d = 0.2539 \pm 0.0079$} is the ratio of fragmentation fractions at $\sqrt{s} = 13$\tev as measured inside the LHCb acceptance~\cite{LHCb-PAPER-2020-046}. The relative efficiency $\xi(\Bs)$ is assumed to be unity, with a 5\% relative systematic uncertainty assigned to account for potential variation in efficiency due to mass and lifetime differences.

\subsection{Fit components}

\begin{figure}[!h]
  \begin{center}
   \includegraphics[width=0.9\linewidth]{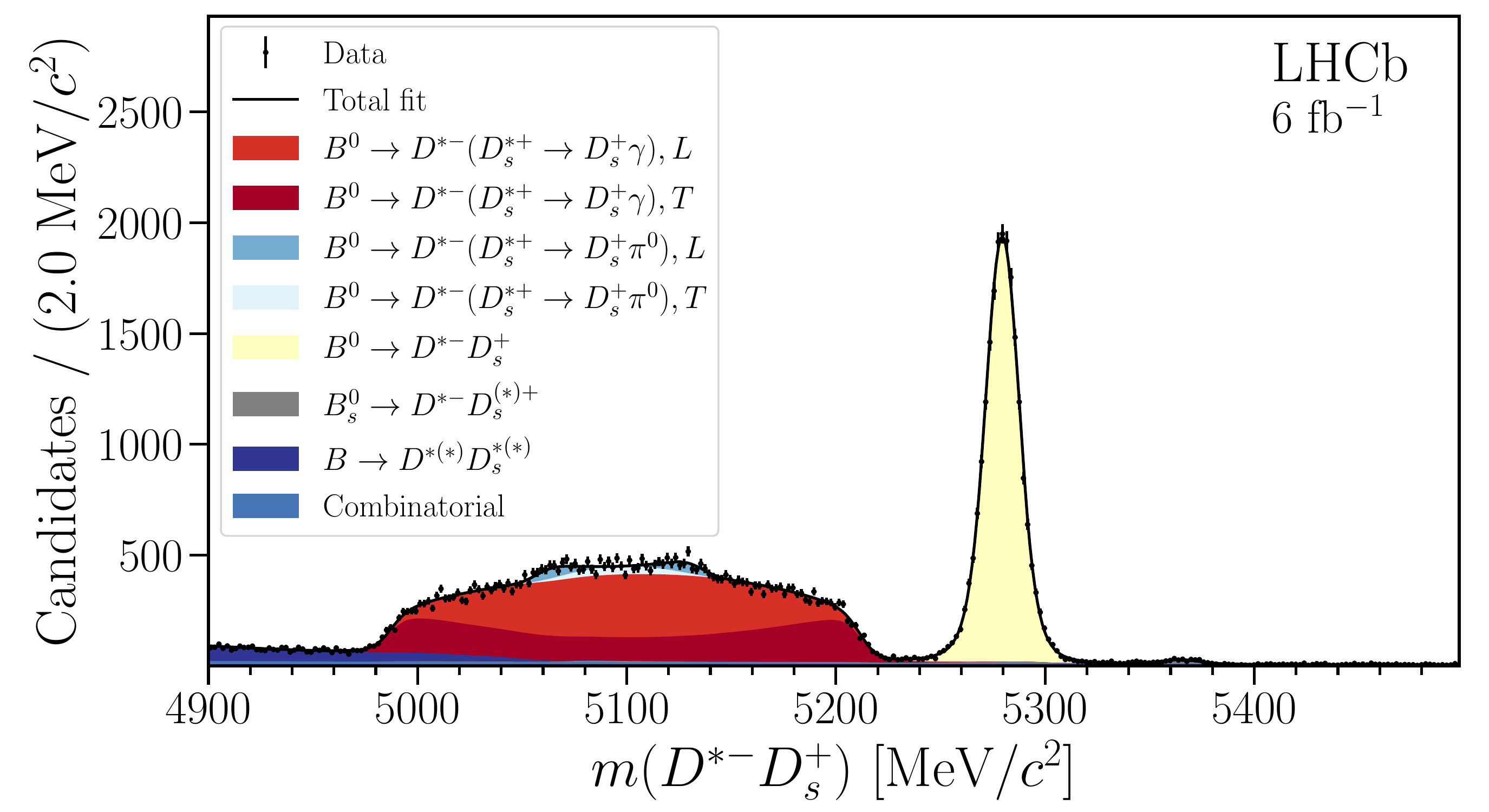}\\
   \includegraphics[width=0.9\linewidth]{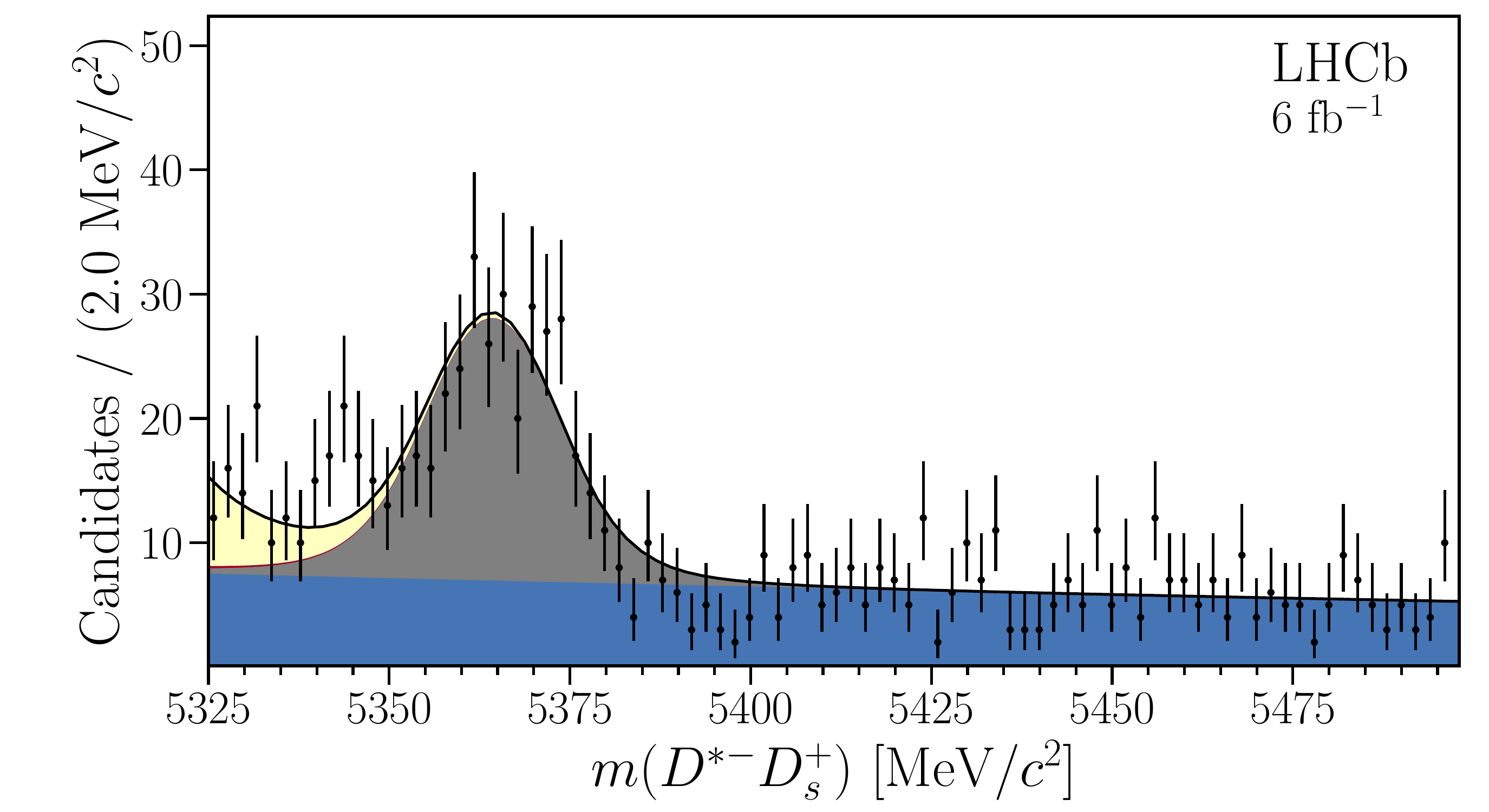}
   \end{center}
   \caption{(Top) Distribution of $m(D^{*-}D_s^+)$ for selected candidates in data, with the fit overlaid. Where indicated, $L$ ($T$) represents longitudinally (transverse) polarised decays. (Bottom) Restricted to region for candidates with \mbox{$m(D^{*-}D_s^+) > 5325$\mevcc}, where the Cabibbo-suppressed $B_s^0 \to D^{*-} D_s^+$ contribution is visible. 
  \label{fig:fL_data_fit}}
\end{figure}

The $m(D^{*-}D_s^+)$ distribution of selected candidates is shown in Fig.~\ref{fig:fL_data_fit}, and is dominated by the narrow signal due to fully reconstructed \BdDstDs decays and a broad structure due to \BdDstDsst decays with missing a photon or \piz from the \Dssp decay.
The distribution is modelled as a sum of several components which are described below.

\subsubsection*{$\boldsymbol{B^0 \to D^{*-}D_s^+}$ decays}

Fully reconstructed \BdDstDs decays are modelled using the sum of two Crystal Ball PDFs~\cite{Skwarnicki:1986xj} with a freely varying common mean and width, and a relative yield fraction that is Gaussian-constrained according to simulation. The component PDF tails are modelled on opposite sides, and the tail parameters are Gaussian-constrained from simulation. The branching fraction ratio $\mathcal{R}$ is measured directly in the fit, such that the yield of the $B^0 \to D^{*-}D_s^+$ component is related to the yield of the $B^0 \to D^{*-} (D_s^{*+} \to D_s^+ \gamma)$ component via a freely varying parameter $\mathcal{R}$ and the fixed relative efficiency ratio $\xi$.

\subsubsection*{$\boldsymbol{B^0 \to D^{*-} (D_s^{*+} \to D_s^+ \gamma)}$ decays}

The partially reconstructed $B^0 \to D^{*-} (D_s^{*+} \to D_s^+ \gamma)$ signal is modelled using the sum of a longitudinal component and a transverse component, where a freely varying parameter \fL determines the relative proportion of the longitudinal component. To derive invariant-mass PDFs for each component, fits are performed to simulated samples of pure longitudinal and transverse polarised decays as shown in Appendix~\ref{app:m_DstDs_costheta_X} in Fig.~\ref{fig:RapidSim_fits}. The $m(D^{*-}D_s^+)$ distributions are modelled with parabolas convolved with Gaussian resolution functions, where the parabolas are based on the $\cos\theta_X$ dependence in Eq.~(\ref{eq:costheta_X_decay_rate}). This approach closely follows the method used in Refs.~\cite{LHCb-PAPER-2017-021} and~\cite{LHCb-PAPER-2020-036} for \CP violation studies of partially reconstructed \mbox{$B^- \to D^{*0} h^-$} with \mbox{$D^{*0} \to D \gamma/\pi^0$} decays, where $h^-$ is a pion or a kaon and the neutral particle produced in the $D^{*0}$ decay is not reconstructed. The total yield of the $B^0 \to D^{*-} (D_s^{*+} \to D_s^+ \gamma)$ component, \mbox{$\mathcal{N}(B^0 \to D^{*-} (D_s^{*+} \to D_s^+ \gamma))$}, varies freely and is used along with $\mathcal{R}$ and $\xi$ to set the $B^0 \to D^{*-} D_s^+$ component yield. All PDF parameters for the \mbox{$B^0 \to D^{*-} (D_s^{*+} \to D_s^+ \gamma)$} component are fixed in the data fit, and are varied within their uncertainties to determine the systematic uncertainties on \fL, $\mathcal{R}$, and $r(\Bs)$.

\subsubsection*{$\boldsymbol{B^0 \to D^{*-} (D_s^{*+} \to D_s^+ \piz)}$ decays}

A contribution from $B^0 \to D^{*-} (D_s^{*+} \to D_s^+ \piz)$ decays, where the neutral pion from the $D_s^{*+}$ decay is not reconstructed, is modelled in a similar manner to the \mbox{$B^0 \to D^{*-} (D_s^{*+} \to D_s^+ \gamma)$} signal. The contribution from this mode is small compared to the signal due to the lower branching fraction of the $D_s^{*+} \to D_s^+ \piz$ decay~\cite{PDG2020}. To determine the invariant-mass PDF for this contribution, separate simulated samples of pure longitudinal and transverse decays are fitted with angular functions convolved with Gaussian resolution functions; all shape parameters are fixed in the data fit. The relative proportion of longitudinal and transverse decays is determined by \fL, where \fL is shared with the $B^0 \to D^{*-} (D_s^{*+} \to D_s^+ \gamma)$ signal decay. The yield of this contribution is fixed relative to the \BdDstDsst component using PDG $D_s^{*+}$ branching fractions~\cite{PDG2020}. Both the fixed PDF parameters and branching fractions are varied within their uncertainties to determine the systematic uncertainty on \fL, $\mathcal{R}$, and $r(\Bs)$.

\subsubsection*{Background from higher-excited charm states}

At low $m(D^{*-}D_s^+)$ values, decays involving higher-excited charm states contribute when one or more particles are not reconstructed. To model the effective contribution from this feed-down background, simulated samples of \mbox{$B^0 \to (D_1(2420)^- \to D^{*-} \piz) D_s^+$}, \mbox{$B^0 \to D^{*-} (D_{s1}(2460)^+ \to D_s^+ \gamma)$}, \mbox{$B^0 \to D^{*-} (D_{s1}(2460)^+ \to (D_s^{*+} \to D_s^+ \gamma) \piz)$}, and \mbox{$B^0 \to (D_1(2420)^- \to D^{*-} \piz) (D_s^{*+} \to D_s^+ \gamma)$} decays generated using \textsc{RapidSim} are studied. The $D_1(2420)^-$ modes are taken as a proxy to represent contributions from similar decays involving $D_1(2430)^-$ and $D_2^*(2460)^-$ mesons, and decays involving two higher-excited charm states are expected to be negligibly small. Invariant-mass fits to simulated events in the \mbox{4900--5350\mevcc} region are performed using sums of several parabolas convolved with resolution functions, where all shape parameters are subsequently fixed in the data fit and varied within their uncertainties to determine the systematic uncertainty. Polarised decays involving two vector mesons are generated using the world-average value of \fL in \BdDstDsst decays~\cite{PDG2020}. Alternative samples are generated with a $\pm 20\%$ variation in \fL to evaluate the change in PDF shape parameters, and the differences observed are assigned as a source of systematic uncertainty. The degree of variation introduced in \fL is motivated by comparing the polarisation fractions measured in several $\Bzb \to D \omega$ decays, where \mbox{$D \in \{D^{*0}$, $D_1(2420)^0$, $D_1(2430)^0$, $D_2^*(2460)^0\}$}~\cite{PDG2020}. The yields of each feed-down contribution are Gaussian-constrained relative to the \BdDstDs yield using a product of PDG branching fractions~\cite{PDG2020}, efficiencies for the $m(D^{*-}D_s^+)$ mass window requirement taken from simulation, and a factor of two to account for the similar expected contributions from $B^\pm$ decays. An additional factor of $0.20 \pm 0.04$ is included for the $D_1(2420)^- D_s^{(*)+}$ modes, in order to model the $B \to (\bar{D}^{**} \to D^{*-} X) D_s^{(*)+}$ rate relative to $B \to D^{*-} D_s^{(*)+}$. This factor is motivated by control mode studies of the rate of $B^+ \to D^{*-} D_s^+ \pi^+$ decays relative to $B^0 \to D^{*-} D_s^+$ decays.

\subsubsection*{Combinatorial background}

Background from random track combinations is modelled using an exponential function with a freely varying shape parameter and yield. Due to the application of mass windows for the charm-meson candidates and a $D_s^+$ candidate flight requirement, the combinatorial background is found to be small across the full $m(D^{*-} D_s^+)$ range considered.

\subsubsection*{Contributions from $\Bs$ decays}

The contribution from Cabibbo-suppressed $\Bs \to D^{*-} D_s^+$ decays falls at higher $m(D^{*-}D_s^+)$ values than the $B^0 \to D^{*-} D_s^+$ decay due to the larger mass of the $\Bs$ meson. This decay is modelled using the same PDF parameterisation as the $B^0 \to D^{*-} D_s^+$ peak, but with independent and freely varying mean and width parameters. The branching fraction ratio $r(\Bs)$ varies freely in the fit, such that the $\Bs \to D^{*-} D_s^+$ yield is determined by $r(\Bs)$, $\xi(\Bs)$, and the external value of $f_s/f_d$. 

Partially reconstructed $\Bs \to D^{*-} D_s^{*+}$ decays are modelled using the same parameterisation as that for $B^0 \to D^{*-} D_s^{*+}$ decays, but with an upward shift in mass set using the known $\Bs$--$B^0$ meson mass difference. The rate of this contribution is determined relative to the $B^0 \to D^{*-} D_s^{*+}$ component using the ratio of the $\Bs \to D^{*-} D_s^+$ and $B^0 \to D^{*-} D_s^+$ component yields, with an additional Gaussian-constrained factor of $1.00 \pm 0.33$ included to allow for potential differences between the $\Bs \to D^{*-} D_s^{*+}$ and $\Bs \to D^{*-} D_s^{+}$ decay rates over a range 0--2. The longitudinal polarisation fraction of the $\Bs \to D^{*-} D_s^{*+}$ component is Gaussian constrained to the value $0.52 \pm 0.16$ based on the world average value for $B^0 \to D^{*-} D_s^{*+}$ decays~\cite{Aubert:2003jj,PDG2020}, where the permitted variation allows for $\fL$ values in the range 0--1.


\subsection{Results}

The fit to the $m(D^{*-}D_s^+)$ distribution in data is shown in Fig.~\ref{fig:fL_data_fit}, where candidates with $m(D^{*-} D_s^+) > 5325$\mevcc are shown on a separate $y$-axis scale in order to highlight the \mbox{$\Bs \to D^{*-} D_s^+$} peak.  Yields of \mbox{$\mathcal{N}(B^0 \to D^{*-} (D_s^{*+} \to D_s^+ \gamma)) = \ndstdsststat$}, \mbox{$\mathcal{N}(B^0 \to D^{*-} D_s^+) = \ndstdsstat$}, and \mbox{$\mathcal{N}(\Bs \to D^{*-} D_s^+) = \nbstodstdsstat$} are obtained, where the uncertainties quoted are statistical only. Studies with pseudoexperiments indicate that the central values and uncertainties of the yields are unbiased. The ratio of branching fractions of \mbox{$B^0 \to D^{*-} (D_s^{*+} \to D_s^+ \gamma)$} decays relative to $B^0 \to D^{*-} D_s^+$ decays is measured to be
\begin{equation*}
\mathcal{R} = \BFratio, 
\end{equation*}
where the first uncertainty is statistical and the second is systematic. In addition, the ratio of branching fractions of the Cabibbo-suppressed \mbox{$B_s^0 \to D^{*-} D_s^+$} decay relative to the \mbox{$B^0 \to D^{*-} D_s^+$} decay is measured to be
\begin{equation*}
r(\Bs) = \BsBFratio, 
\end{equation*}
where the first uncertainty is statistical, the second is systematic, and the third is due to the use of an external value of $f_s/f_d$~\cite{LHCb-PAPER-2020-046}. The systematic uncertainties on $\mathcal{R}$ and $r(\Bs)$ are due to the use of fixed PDF shape parameters and branching fractions in the fit, as well as the use of the relative efficiency corrections $\xi$ and $\xi(\Bs)$. The contributing systematic uncertainties on both branching fraction ratios are summarised in Table~\ref{tab:fL_systs} in Sect.~\ref{sec:Syst}. The value of $\mathcal{R}$ is in agreement with the world average, $\mathcal{R} = 2.07 \pm 0.33$, but has a considerably smaller uncertainty. The measurement of $r(\Bs)$ is a world first, and constitutes the first observation of the Cabibbo-suppressed $\Bs \to D^{*-} D_s^+$ decay with a statistical significance of seven standard deviations. The significance is calculated by determining the difference in $r(\Bs)$ from zero, where both the statistical and systematic uncertainties are considered.

The longitudinal polarisation fraction in $B^0 \to D^{*-} (D_s^{*+} \to D_s^+ \gamma)$ decays is measured to be
\begin{equation*}
\fL = \fLsyst,
\end{equation*}
where the first uncertainty is statistical and the second is systematic. The systematic uncertainty quoted is due to the limited knowledge of the fixed terms used in the fit. This result is in agreement with, but substantially more precise than, the current world-average value. Pseudoexperiment studies indicate that the fitted central value and uncertainty of \fL are unbiased. In the subsequent analysis of fully reconstructed \BdDstDsst decays presented herein, \fL is fixed to the value measured in the $m(D^{*-}D_s^+)$ fit. This enables the angular acceptance functions for $\cos\theta_D$ and $\cos\theta_X$ to be derived directly from data (see Sect.~\ref{sec:Acc}), rather than modelling such effects using simulation. The $\cos\theta_X$ distribution in particular is sensitive to mis-modelling in the simulation, due to its dependence on the soft photon kinematics which can be distorted by the hardware trigger emulation in the simulation. 

\section{Invariant-mass fit to $\boldsymbol{B^0 \to D^{*-} D_s^{*+}}$ decays}
\label{sec:sFit}

\begin{figure}[!h]
  \begin{center}
  \vspace{0.5cm}
   \includegraphics[width=0.9\linewidth]{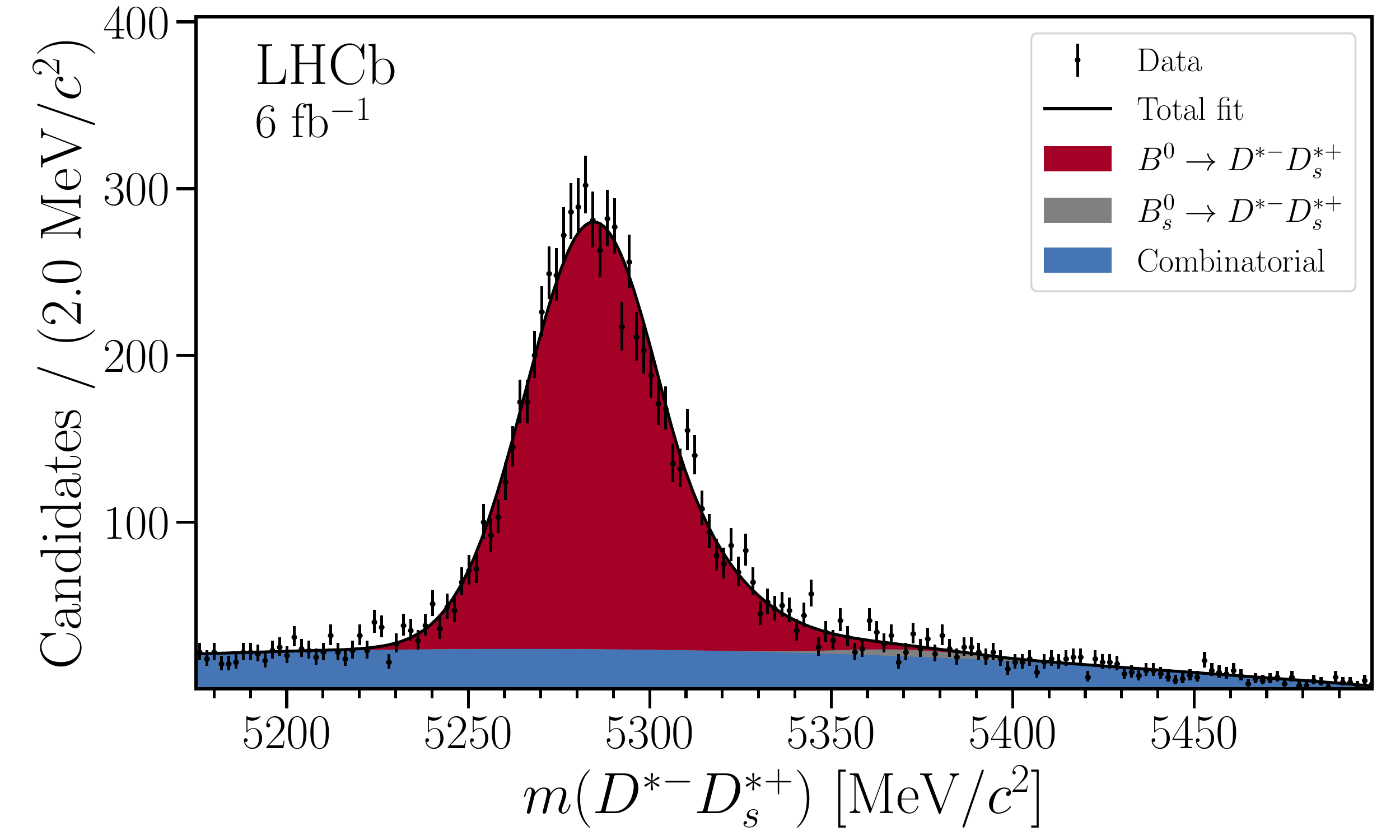}
   \end{center}
   \caption{Distribution of $m(D^{*-}D_s^{*+})$ for selected candidates in data, with the fit overlaid.
  \label{fig:mB_sWeight_fit}}
\end{figure}

To derive signal weights for the angular analysis, a binned maximum-likelihood fit to the $m(D^{*-}D_s^{*+})$ distribution of fully reconstructed \BdDstDsst candidates in data is performed using the \sPlot method~\cite{Pivk:2004ty}. The $m(D^{*-}D_s^{*+})$ distribution is shown in Fig.~\ref{fig:mB_sWeight_fit}, where the fit is overlaid. This fully reconstructed sample contains 17\% of the candidates used in the $m(D^{*-} D_s^+)$ fit in Sect.~\ref{sec:fL_Fit}; the smaller sample size is attributed to the limited soft photon reconstruction efficiency, and the application of additional requirements on the photon and $D_s^{*+}$ candidate. Potential background contributions from \BdDstDsst decays with $D_s^{*+} \to D_s^+ \pi^0$ are determined to be negligible within the window of $D_s^{*+}$--$D_s^+$ mass difference considered.

The fit is performed using the sum of a $B^0 \to D^{*-} (D_s^{*+} \to D_s^+ \gamma)$ signal component, a \mbox{$\Bs \to D^{*-} (D_s^{*+} \to D_s^+ \gamma)$} background component, and a combinatorial background component. The signal is described using the sum of two Crystal Ball PDFs which share a common freely varying mean and width. The tail parameters and relative fraction of the two Crystal Ball PDFs are constrained from fits to simulation, where the component PDFs are required to have tails on opposite sides. The yield of the signal component varies freely, and is found to be $\ndstdsstfullreco$. The \mbox{$\Bs \to D^{*-} (D_s^{*+} \to D_s^+ \gamma)$} component is modelled using the same PDF as the signal, but with a mean shifted upwards using the known $\Bs$--$B^0$ meson mass difference. The rate of this contribution is fixed relative to signal using the proportions determined in the $m(D^{*-} D_s^+)$ fit. The combinatorial background is modelled using a second-order Chebyshev polynomial, where the yield and shape parameters of this contribution vary freely. As the background distribution is not known a priori, an alternative parameterisation using a Gaussian function is also used to model the combinatorial background. The signal weights derived from this alternative model are used to determine the systematic uncertainty on the helicity observables. A Gaussian function is used as it provides a background description of equivalent quality to the second-order Chebyshev, whereas linear and exponential background models do not describe the background sufficiently well.

An underlying assumption of the \sPlot method used to derive per-candidate signal weights is that the discriminating variable, in this case $m(D^{*-}D_s^{*+})$, is uncorrelated with the target distributions to be studied with weights applied, in this instance the decay angles. Due to a common underlying dependence on the decay product kinematics, the invariant mass and decay angles do exhibit some degree of correlation. To assess the potential bias from this, a combined four-dimensional simulated sample of signal and background events is generated in $m(D^{*-}D_s^{*+})$ and the decay angles. The background sample is generated according to the $m(D^{*-}D_s^{*+})$ background shape observed in data, and with a flat distribution in each of the decay angles. The total simulated sample contains the same number of signal and background events as measured in the $m(D^{*-}D_s^{*+})$ data fit. A fit to the $m(D^{*-}D_s^{*+})$ distribution of the simulated sample is performed to derive signal weights, which are then applied when creating histograms in the decay angles. Using $\chi^2$ tests, these histograms are compared to histograms of the decay angles created using only the simulated signal sample. All of the signal-weighted distributions are found to agree with the pure signal distributions, indicating that no significant biases are incurred from the use of signal weights. 

\FloatBarrier 

\section{Angular acceptance functions}
\label{sec:Acc}

Due to experimental acceptance and resolution effects, the angular distributions in data are distorted relative to the true distributions. As the decay angles are measured with a relative resolution of 2--4\% according to simulation, the dominant effect on the experimental angular distributions is due to the acceptance. This effect must be modelled in the angular fit in order to derive unbiased measurements of the helicity observables, which is achieved by multiplying the true differential decay rate PDF by acceptance functions defined in each of the decay angles. This approach assumes that the total angular acceptance can be factorised into a product of the individual acceptance functions for each angle, which is validated using a simulated sample of signal decays generated according to Eq.~(\ref{eq:decay_rate}) with the world-average value of $\fL = 0.52$~\cite{Aubert:2003jj,PDG2020}, \mbox{$|H_+| = |H_-| = \sqrt{(1 - \fL)/2}$}, and all phases equal to zero. The efficiency of a cut applied to all three decay angles together, $\epsilon_{xyz}$, is compared with a product of the efficiencies for cuts applied separately to each decay angle, $\epsilon' = \epsilon_x \times \epsilon_y \times \epsilon_z$; the values of $\epsilon_{xyz}$ and $\epsilon'$ are found to agree within the uncertainties due to the use of finite simulated samples.

\subsection{Acceptance functions for \boldmath{$\cos\theta_D$} and \boldmath{$\cos\theta_X$}}

\begin{figure}[!h]
  \begin{center}
   \includegraphics[width=0.49\linewidth]{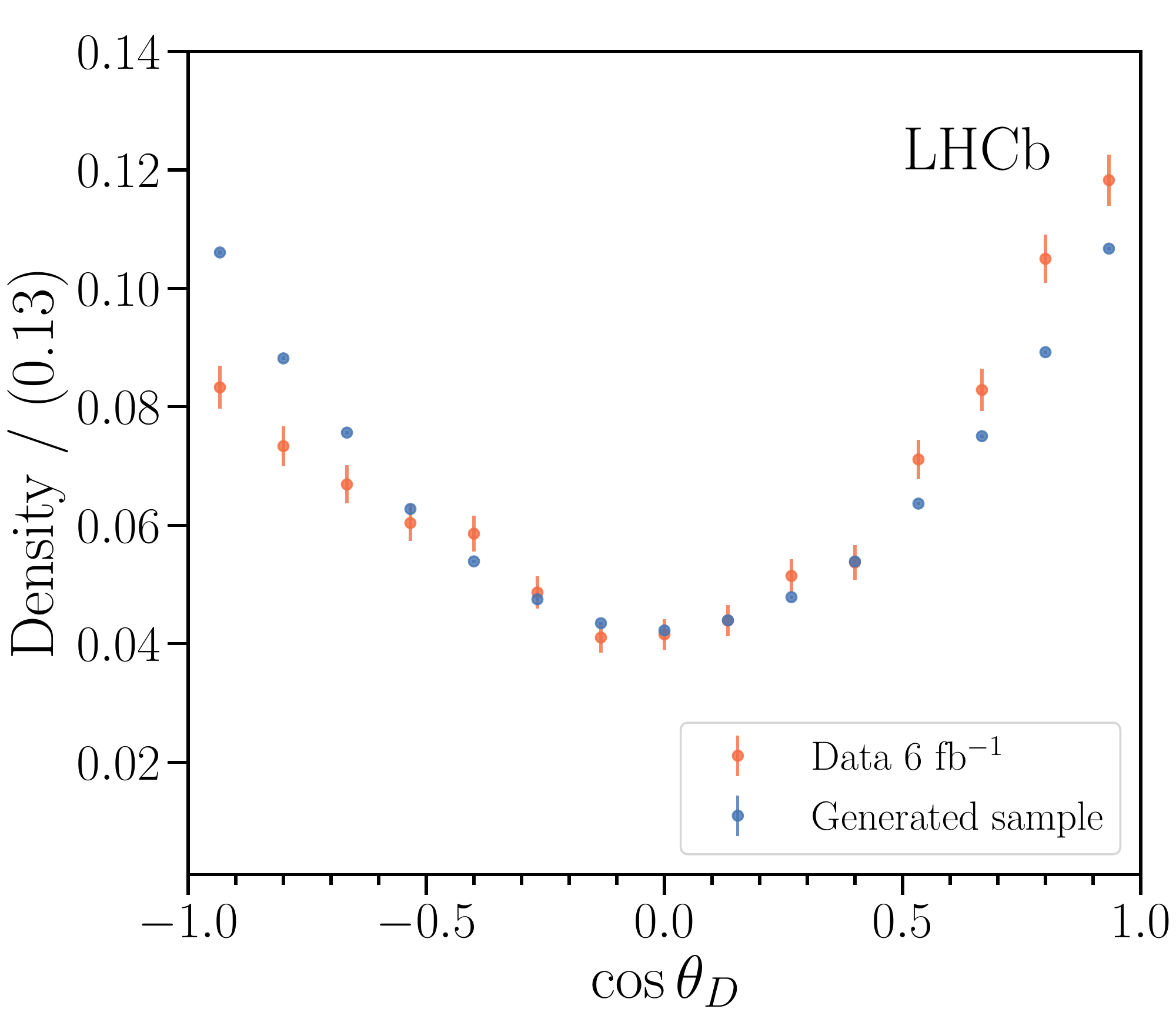}
   \includegraphics[width=0.49\linewidth]{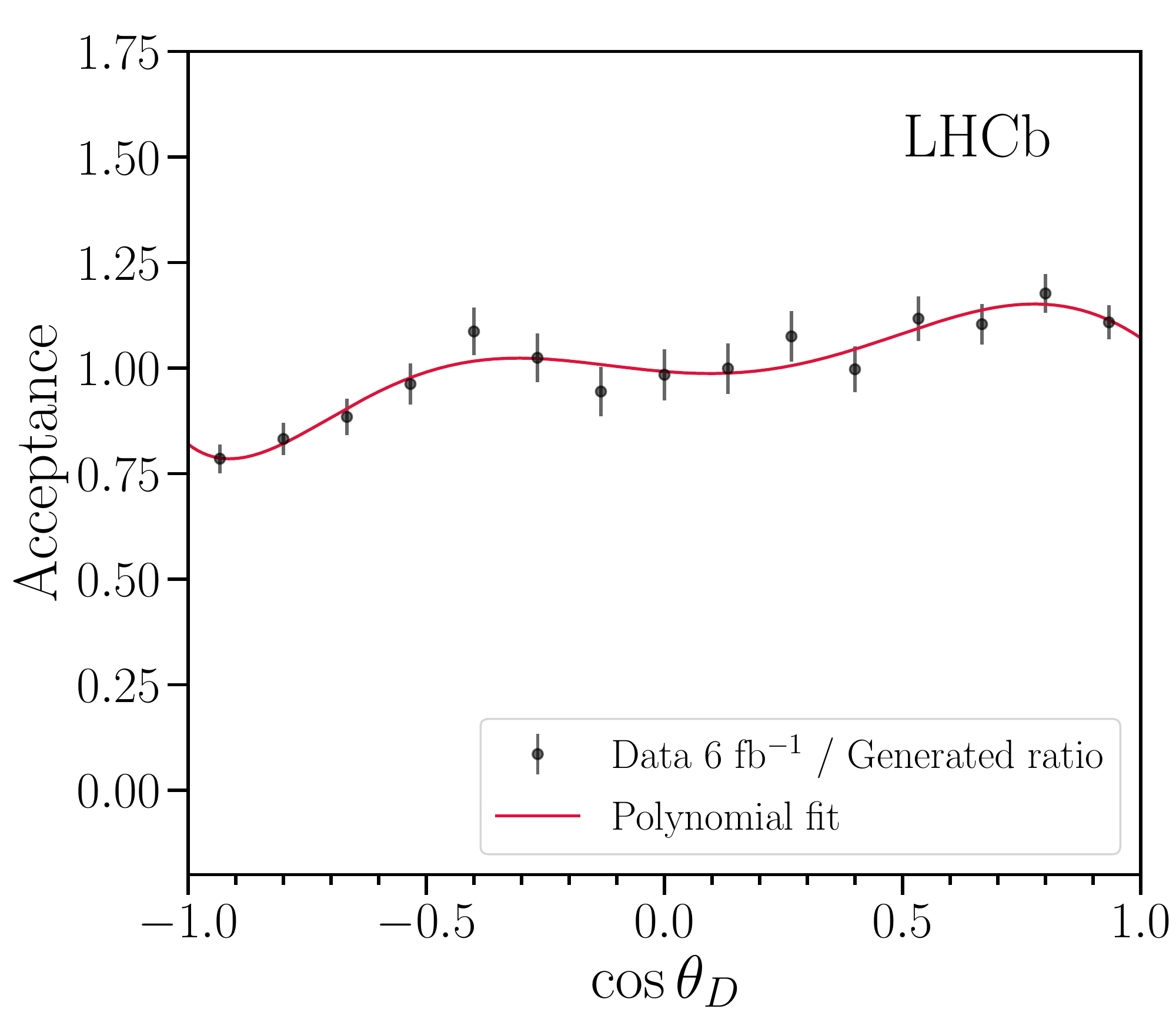}\\
   \includegraphics[width=0.49\linewidth]{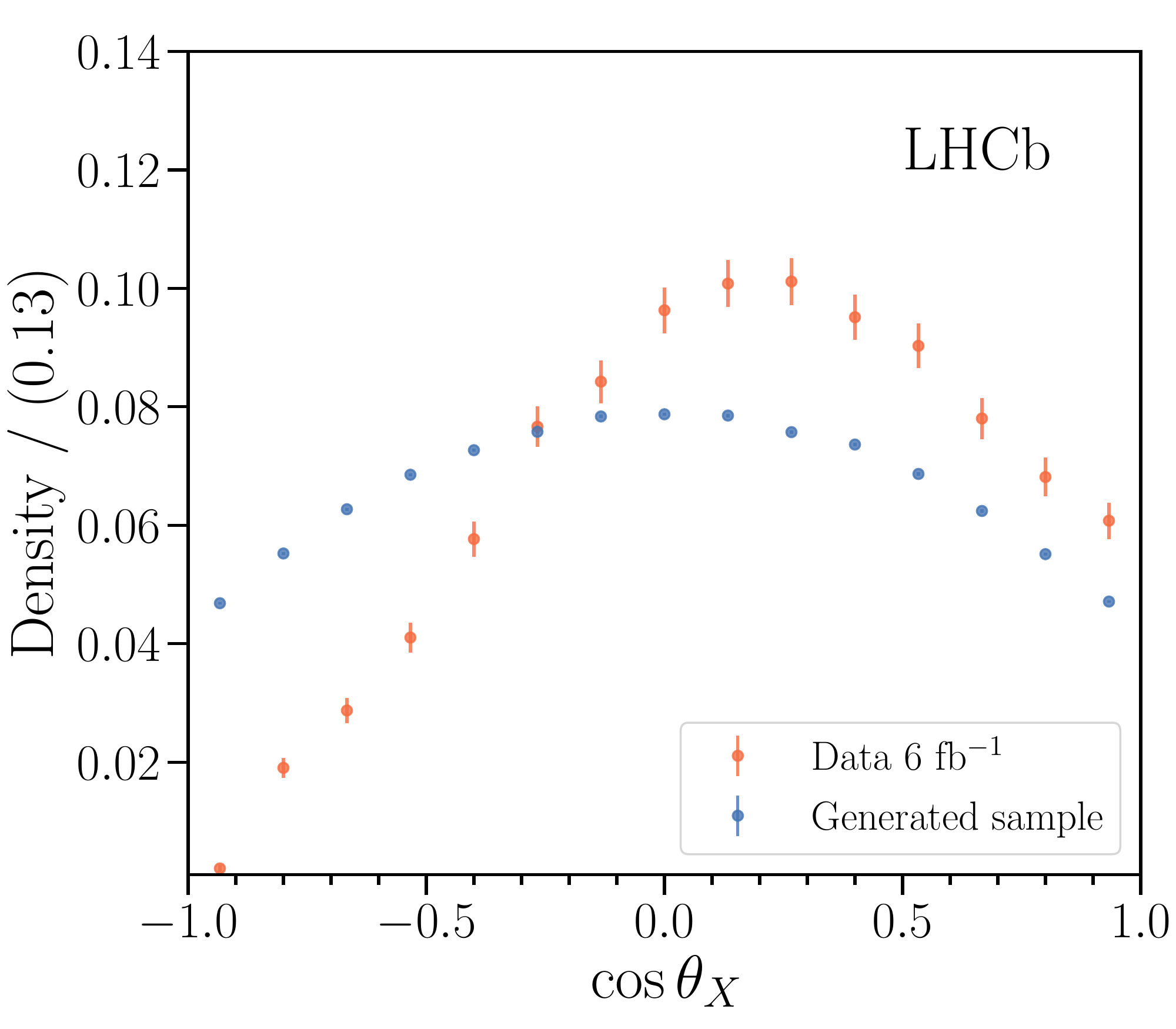}
   \includegraphics[width=0.49\linewidth]{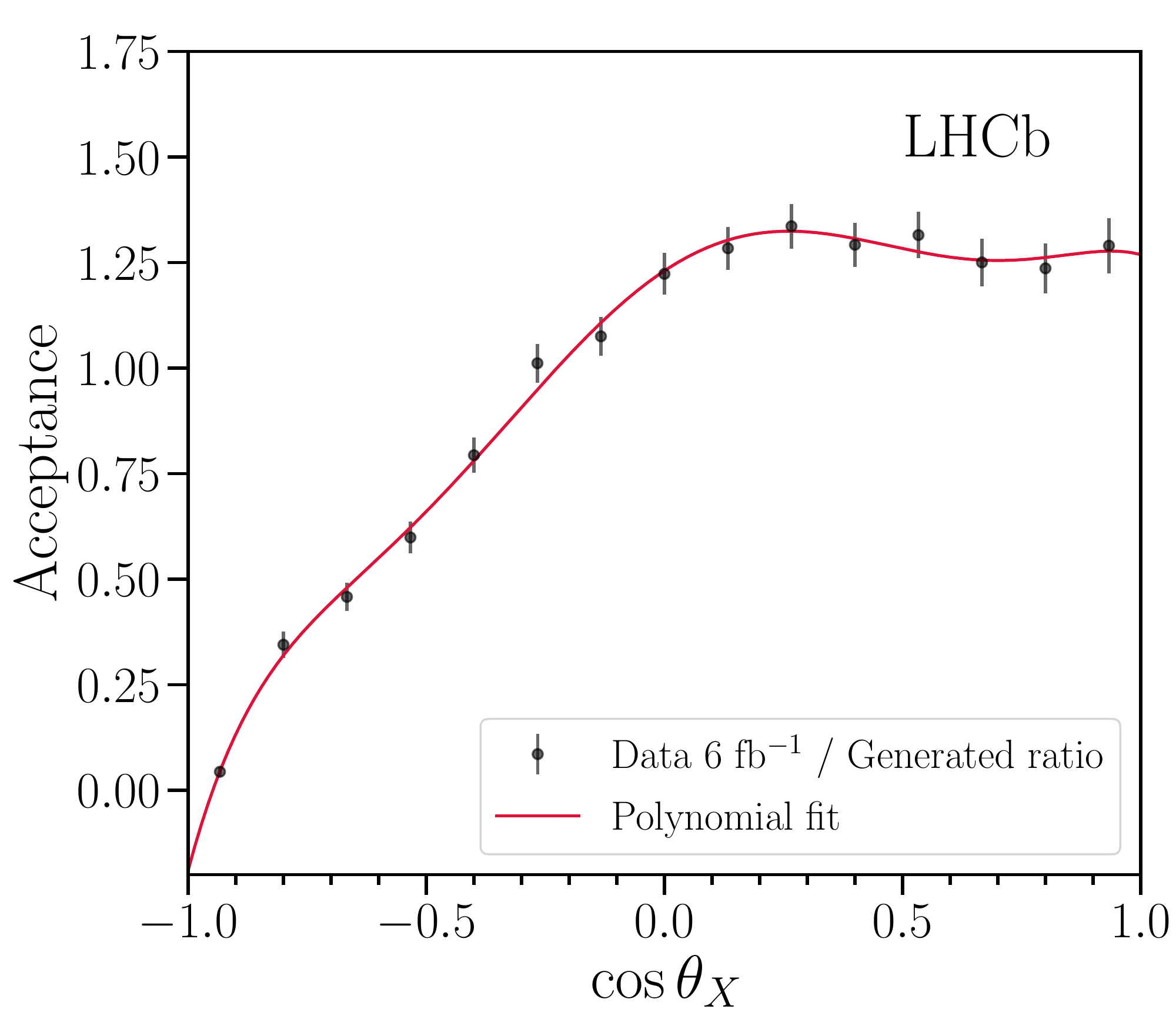}\\
   \end{center}
   \caption{(Left) Comparison of signal-weighted data and generated (top) $\cos\theta_D$ and (bottom) $\cos\theta_X$ distributions, where the differences observed are due to the experimental acceptance and resolution. (Right) Data to generated sample ratios, with the polynomial fits overlaid.
  \label{fig:costheta_D_X_acceptance}}
\end{figure}

The acceptance functions for $\cos\theta_D$ and $\cos\theta_X$ are derived from data. Binned normalised distributions in each decay angle are produced by creating histograms of the fully reconstructed \BdDstDsst candidates in data with signal weights applied. The only physical observable that can alter the shape of the one-dimensional $\cos\theta_D$ and $\cos\theta_X$ distributions is \fL, which is known from the $m(D^{*-}D_s^+)$ fit. The acceptance is thus determined by comparing the data distributions with angular distributions generated with \textsc{RapidSim} using the value of \fL measured in Sect.~\ref{sec:fL_Fit}.
In the generated sample, no detector acceptance or resolution effects are included. The signal-weighted data and generated signal distributions in $\cos\theta_D$ and $\cos\theta_X$ are compared in Fig.~\ref{fig:costheta_D_X_acceptance} (left column). The observed differences between data and the generated sample are attributed to the experimental acceptance and resolution, since both distributions share a common \fL value. The $\cos\theta_X$ distribution in particular exhibits substantial acceptance effects, where candidates at low $\cos\theta_X$ are preferentially removed. This warping is due to the application of photon $\pt$ requirements in the selection, which bias the sample to more positive values of $\cos\theta_X$.

To determine acceptance functions for $\cos\theta_D$ and $\cos\theta_X$, the binned ratios of data to the generated sample are fitted with sixth-order polynomial functions. The fits are shown in Fig.~\ref{fig:costheta_D_X_acceptance} (right column), and the polynomial coefficients are employed as fixed terms in the angular fit in Sect.~\ref{sec:Fit}. To determine the systematic uncertainty on the helicity observables due to the finite dataset used in the acceptance fits, the acceptance function coefficients are varied within their uncertainties according to the acceptance fit covariance matrices. When determining the systematic uncertainty due to the use of a fixed \fL value in the angular analysis, the acceptance fits are performed many times with \fL varied randomly within its total measured uncertainty. 

For values of $\cos\theta_X$ close to $-1$, which correspond to the smallest photon momentum values, the acceptance function becomes slightly negative due to limited data statistics in this region. A fiducial cut of $\cos\theta_X > -0.9$ is applied to data in order to remove the region of negative modelled acceptance; this requirement is found to have a negligible impact on the measured helicity observables. 

\subsection{Acceptance function for \boldmath{$\chi$}}

\begin{figure}[!h]
  \begin{center}
   \includegraphics[width=0.49\linewidth]{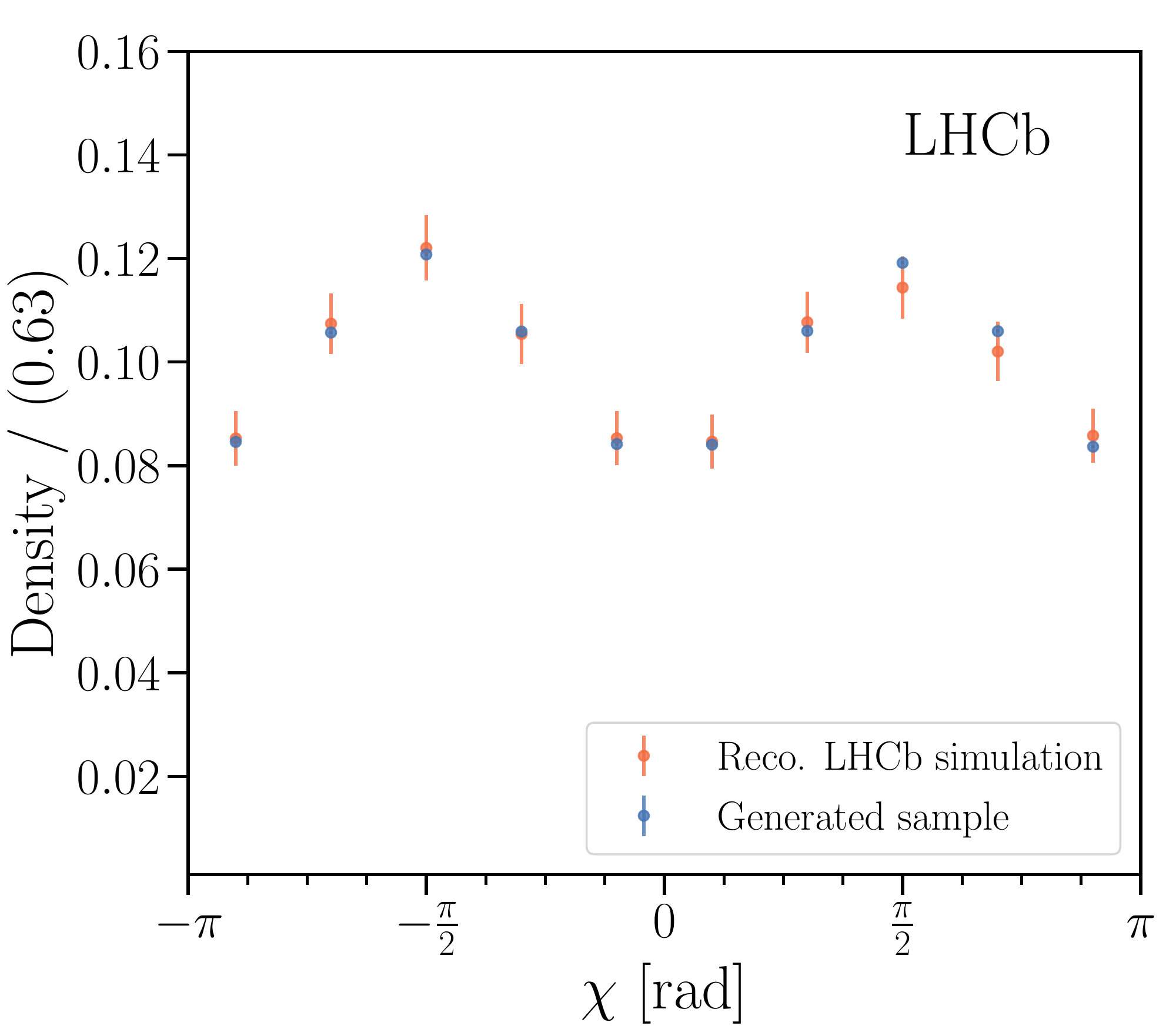}
   \includegraphics[width=0.49\linewidth]{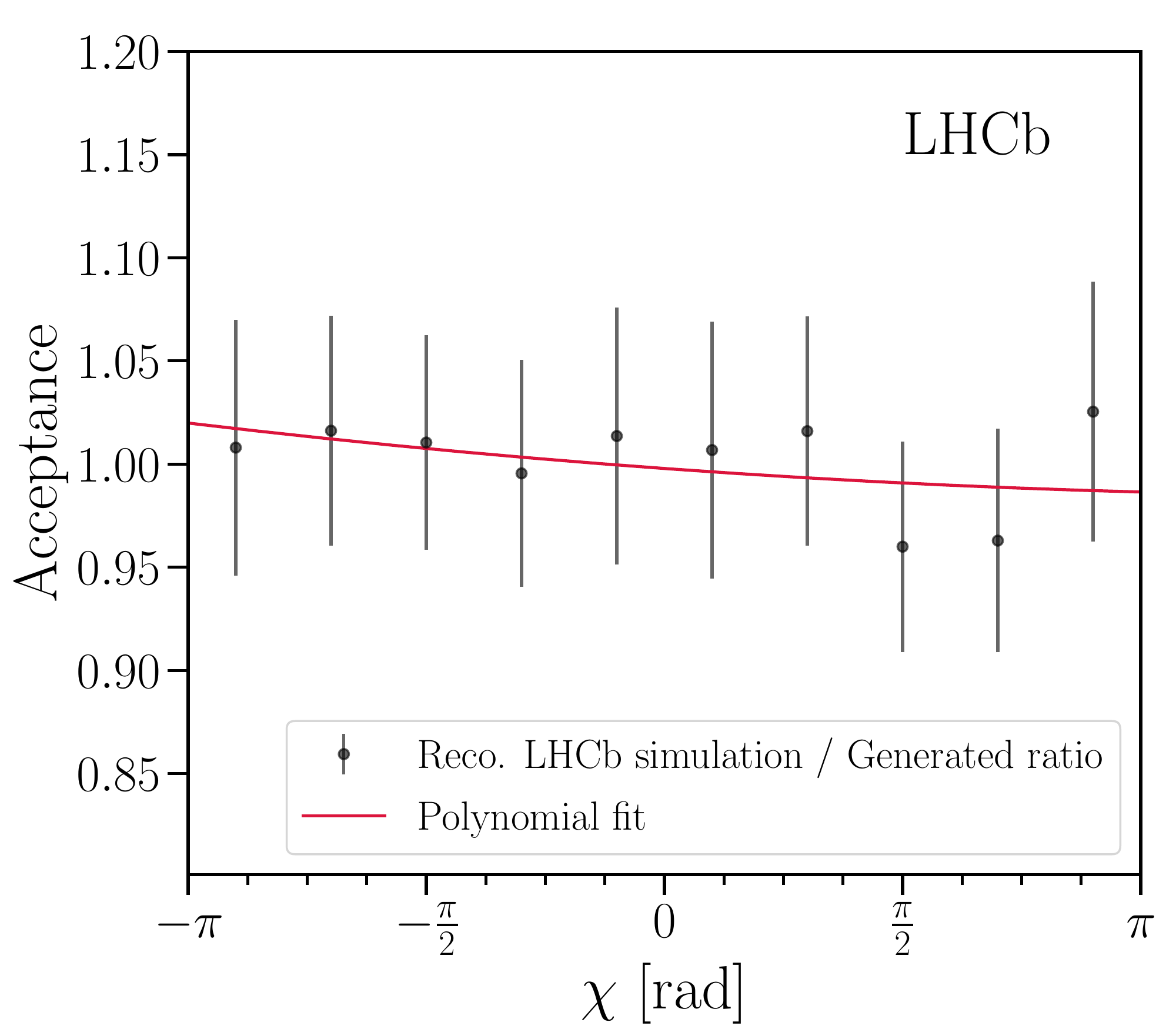}
   \end{center}
   \caption{Comparison of reconstructed $\chi$ distribution in a fully-simulated \BdDstDsst sample and the generated $\chi$ distribution in a \textsc{RapidSim} sample produced with the same helicity amplitude model (left). The ratio is fitted with a second-order polynomial to determine the acceptance function for use in the data fit (right). 
  \label{fig:chi_acceptance}}
\end{figure}

In Eq.~(\ref{eq:decay_rate}), all of the angular terms that are sensitive to the relative magnitudes and phases of the transverse amplitudes have a dependence on the angle $\chi$. As such, no information on the $\chi$ acceptance can be derived from data. To determine the $\chi$ acceptance, the reconstructed $\chi$ distribution in a sample of fully-simulated \BdDstDsst decays passing all selection requirements is compared to a generated $\chi$ distribution produced using \textsc{RapidSim} with the same model parameters but no acceptance or resolution effects. For this comparison, the simulated samples are generated with the \fL value measured in Sect.~\ref{sec:fL_Fit}, with $|H_-| = |H_+|$ and $\phi_+ = \phi_- = 0$. The binned $\chi$ distributions are shown in Fig.~\ref{fig:chi_acceptance} (left), where good agreement between the reconstructed and generated distributions is found. This indicates that the reconstructed $\chi$ distribution is not strongly modified by acceptance effects. To model residual acceptance effects, the reconstructed to generated $\chi$ ratio is fitted with a second-order polynomial, as shown in Fig.~\ref{fig:chi_acceptance} (right). This function is employed as a fixed correction PDF in the angular fit, and the polynomial coefficients are varied within their uncertainties to determine the systematic uncertainties on the helicity parameters. In this procedure, the correlations between the polynomial coefficients are accounted for using the acceptance fit covariance matrix.

\section{Angular fit to data}
\label{sec:Fit}

\begin{figure}[!h]
  \begin{center}
   \includegraphics[width=0.49\linewidth]{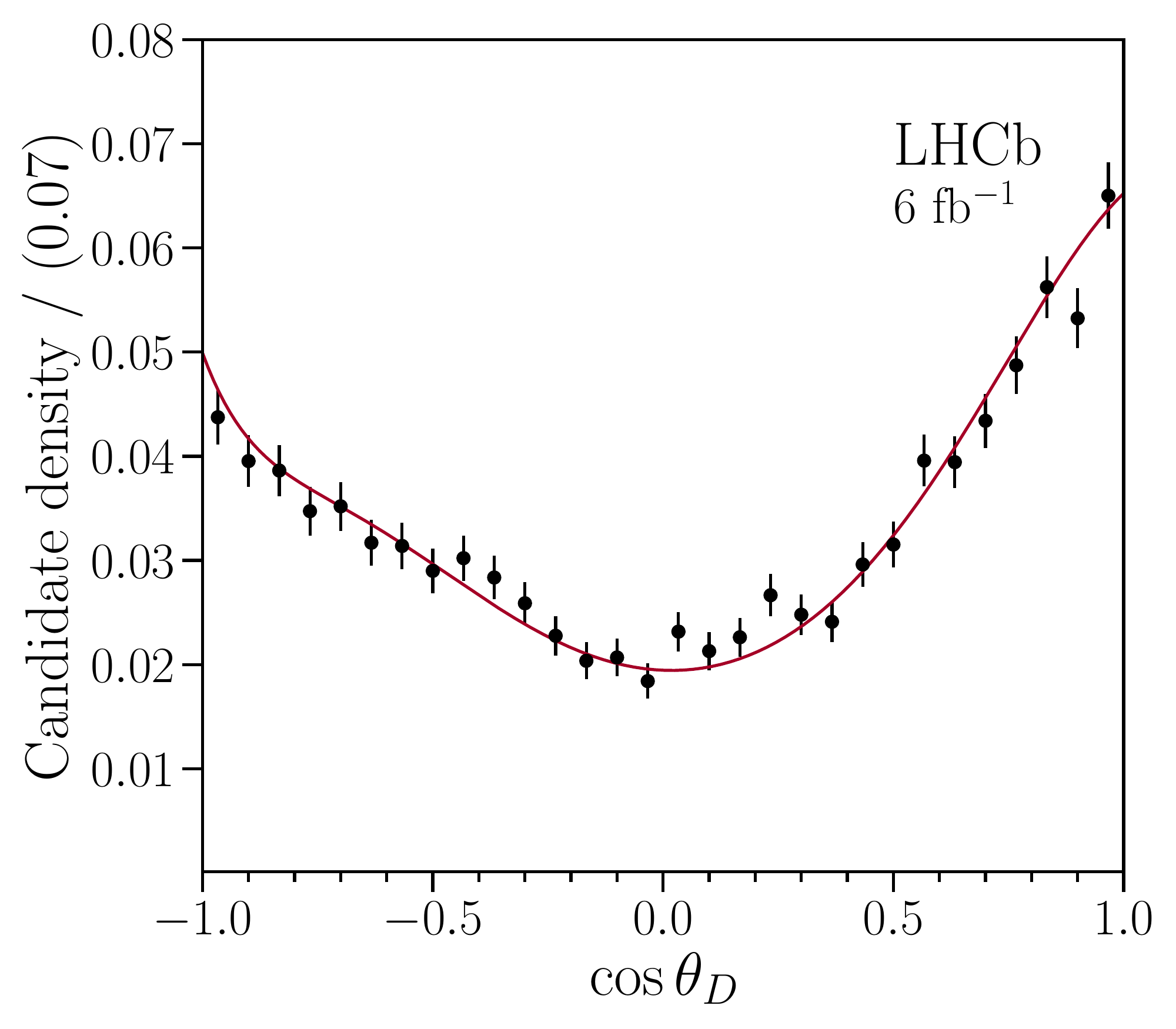}
   \includegraphics[width=0.49\linewidth]{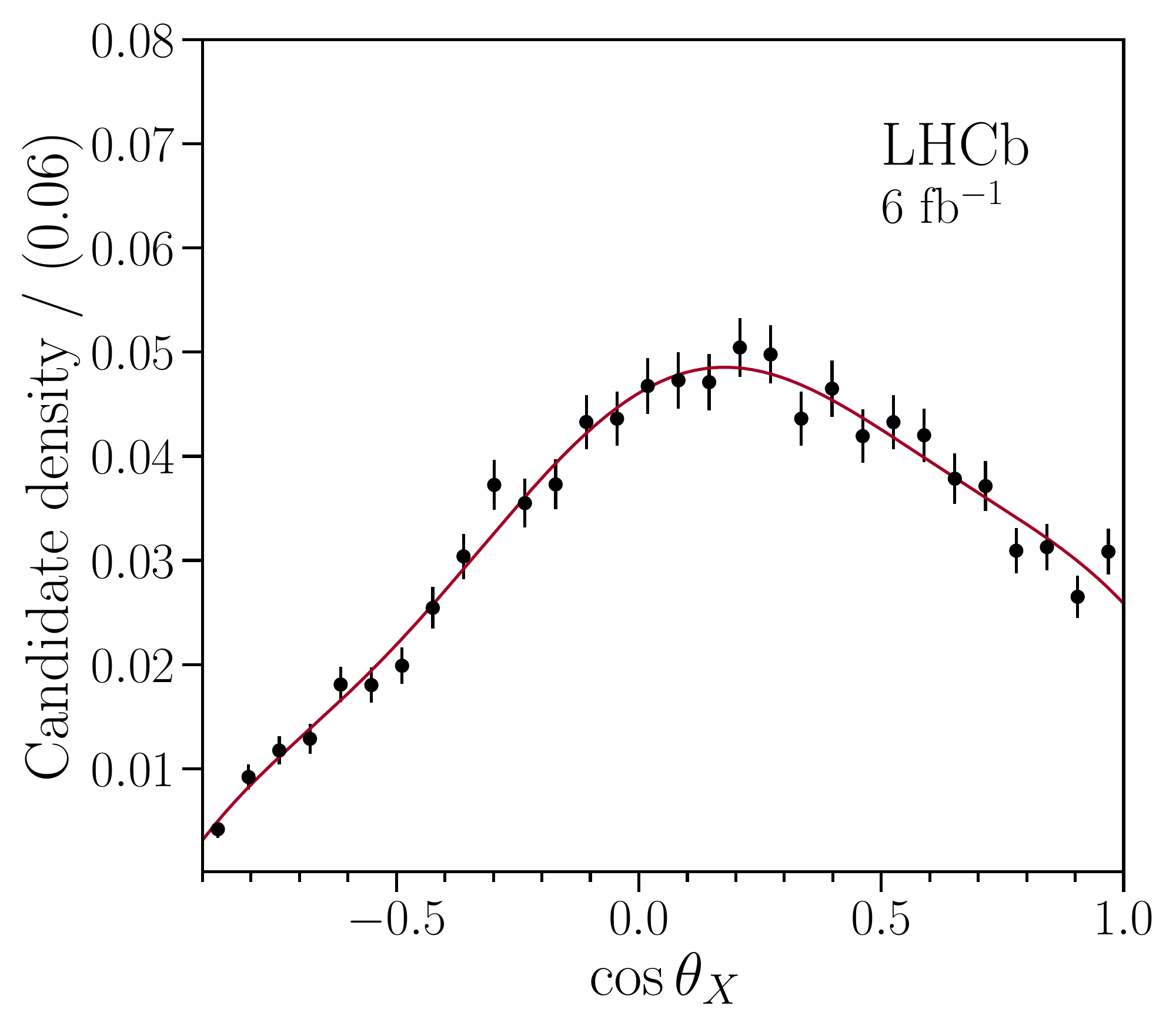} \\
   \includegraphics[width=0.49\linewidth]{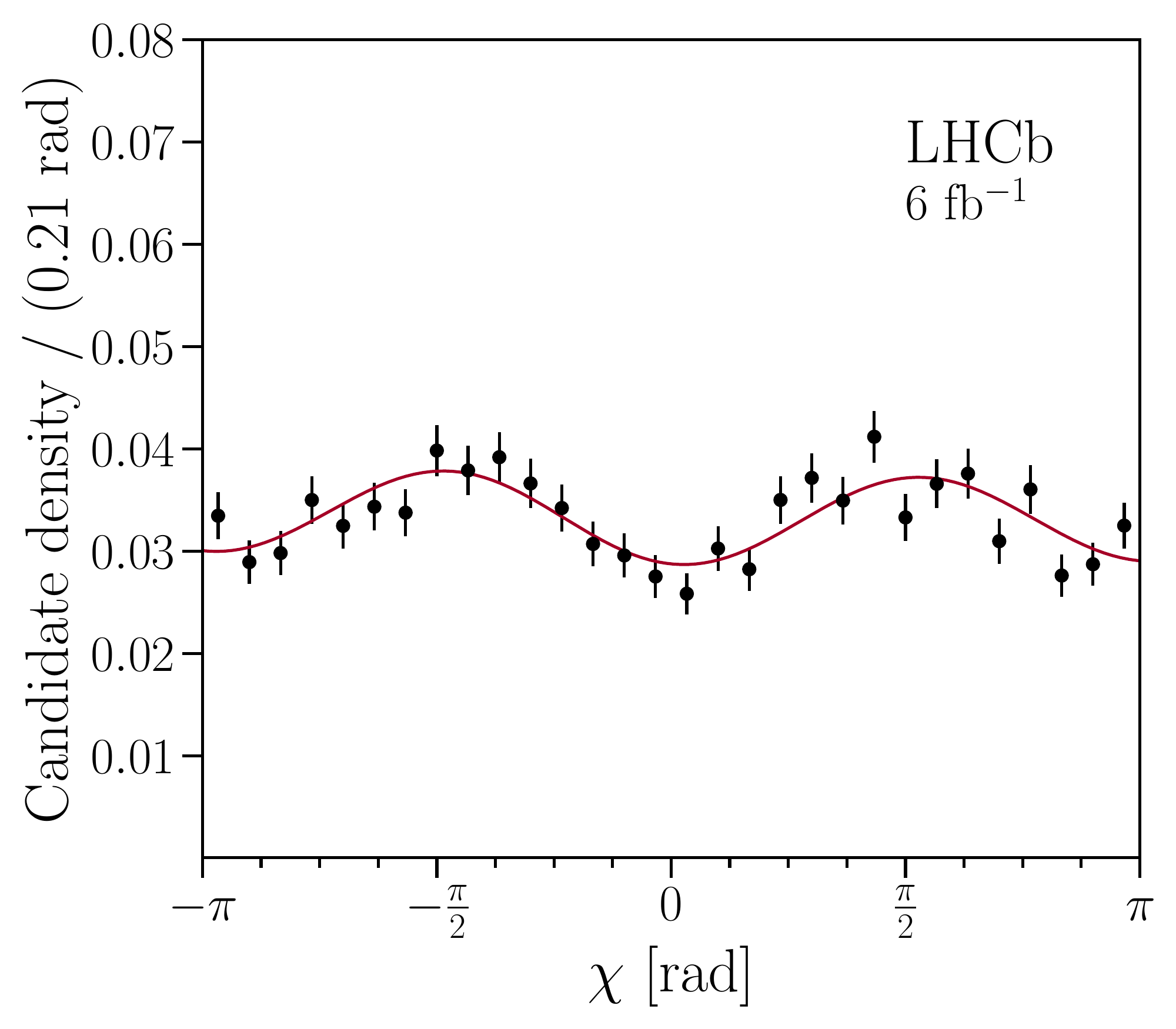}
   \end{center}
   \caption{Decay-angle distributions of signal-weighted \BdDstDsst candidates in data, with the one-dimensional angular fit projections overlaid.
  \label{fig:1D_proj}}
\end{figure}

To measure $|H_-|$, $\phi_-$, and $\phi_+$, an unbinned maximum-likelihood fit to the three-dimensional angular distribution of signal-weighted data is performed using \textsc{zfit}~\cite{Eschle:2019jmu}. For the fit, the $B^0 \to D^{*-} (D_s^{*+} \to D_s^+ \gamma)$ candidates from the $m(D^{*-}D_s^{*+})$ fit in Sect.~\ref{sec:sFit} are used with per-candidate signal weights assigned. The longitudinal polarisation amplitude, $H_0$, is assigned a fixed magnitude $|H_0|$ using the value of \fL measured in Sect.~\ref{sec:fL_Fit}, and its phase is set to the arbitrary value $\phi_0 = 0$. The parameter $|H_+|$ is fully determined by the normalisation of the helicity amplitudes to unity. The signal density at each point in angular phase space is described using Eq.~(\ref{eq:decay_rate}) multiplied by acceptance functions in each of the decay angles. To determine the statistical uncertainties of the observables, the fit applies an asymptotic correction to the covariance matrix as detailed in Ref.~\cite{Langenbruch:2019nwe}, which correctly accounts for the use of signal-weighted data. The distributions for each decay angle are shown in Fig.~\ref{fig:1D_proj}, with the one-dimensional fit projections overlaid.

Studies with pseudoexperiments are performed to determine the level of bias present in the results, where pull distributions of mean $\mu_P^x$ and width $\sigma_P^x$ are constructed for each observable $x$. The pull distributions for each helicity observable are found to follow Gaussian distributions closely, where $\sigma_P^{|H_-|}$ is consistent with unity. However, $\sigma_P^{\phi_+} = 1.14 \pm 0.02$ and $\sigma_P^{\phi_-} = 1.12 \pm 0.02$, indicating that the default fit uncertainties for these observables are underestimated. The mean values of the pulls for the transverse phases are consistent with zero, but $\mu_P^{|H_-|} = -0.14 \pm 0.02$. These biases are traced to the finite size of the fitted dataset, and are found to resolve when pseudoexperiment datasets containing more events than are present in data are generated. The values of $\mu_P^x$ and $\sigma_P^x$ are used to correct the default fit results $x \pm \sigma_x$ as follows
\begin{align}
    x^c &= x - \mu_P^x \times \sigma_x \\
    \sigma_x^c &= \sigma_P^x \times \sigma_x
\end{align}
where $x^c \pm \sigma_x^c$ are the corrected fit results. In Sect.~\ref{sec:Results}, the results for $|H_-|$, $\phi_+$, and $\phi_-$ are quoted after this correction procedure.

\section{Systematic uncertainties}
\label{sec:Syst}

The values of $\mathcal{R}$, $r(\Bs)$, and \fL measured in Sect.~\ref{sec:fL_Fit} are subject to systematic uncertainties due to limited knowledge of the shape parameters, branching fractions, and relative efficiency corrections used in the fit. To determine these systematic uncertainties, the $m(D^{*-}D_s^+)$ fit to data is performed many times with the parameters randomly varied within their prescribed uncertainties according to Gaussian distributions. This procedure is performed separately for the shape parameters, branching fractions, and efficiency corrections, and the total systematic uncertainties calculated as the sum in quadrature of these contributions. The systematic uncertainties are summarised in Table~\ref{tab:fL_systs}. 

The observables $|H_-|$, $\phi_+$, and $\phi_-$ measured in the angular fit are subject to several systematic uncertainties. Firstly, the angular analysis is performed at a fixed value of \fL, which is used as input in the $\cos\theta_D$ and $\cos\theta_X$ acceptance fits and also to set the value of $|H_0|$ in the angular fit. To determine the systematic uncertainty, the angular analysis is repeated many times with \fL varied within its total uncertainty; the standard deviations of the helicity observable results are taken as the systematic uncertainties. In this procedure, the varied \fL value used in the acceptance fits is shared with the angular fit to ensure consistency. A small systematic uncertainty is also assigned for the use of signal-weighted data, where the angular fit is run many times while varying the signal weights within the signal yield uncertainties from the $m(D^{*-}D_s^{*+})$ fit. To determine the systematic uncertainty from the use of finite samples to obtain the acceptance functions, the acceptance coefficients are varied within their uncertainties according to the acceptance fit covariance matrices. Finally, the angular analysis is repeated with an alternative background model in the $m(D^{*-}D_s^{*+})$ fit, and the differences in central value for each helicity observable are assigned as a systematic uncertainty. The contributing systematic uncertainties are summarised in Table~\ref{tab:angular_systs}.

\begin{table}[!h] 
\begin{center} 
\caption{Systematic uncertainties on the branching fraction ratios and $f_L$ as measured in the $m(D^{*-} D_s^+)$ fit. \label{tab:fL_systs}} 
\begin{tabular}{c|c|c|c} 
\toprule 
Systematic uncertainty & $\mathcal{R}$ & $\mathcal{R}(B_s^0)$ & $f_L$\\ \midrule 
Fixed PDF shape parameters & 0.030 & 0.00197 & 0.0074\\ 
Fixed branching fractions & 0.016 & 0.00004 & 0.0080\\ 
Efficiency corrections & 0.062 & 0.00253 & 0.0001\\ 
\midrule 
Total & 0.071 & 0.00320 & 0.0109\\ 
\bottomrule 
\end{tabular} 
\end{center} 
\end{table}

\begin{table}[!h] 
\begin{center} 
\caption{Systematic uncertainties on the helicity parameters measured in the unbinned angular fit. \label{tab:angular_systs}} 
\begin{tabular}{c|c|c|c} 
\toprule 
Systematic uncertainty & $|H_-|$ & $\phi_+$ & $\phi_-$\\ \midrule 
Fixed $f_L$ in angular fit and $\cos(\theta_{X/D})$ acceptance & 0.0005 & 0.0007 & 0.005\\ 
Use of $_s$Weighted data & 0.0003 & 0.0011 & 0.002\\ 
Statistical uncertainty of acceptance functions & 0.0034 & 0.0132 & 0.044\\ 
$m(D^* D_s^*)$ fit background model & 0.0319 & 0.0156 & 0.025\\ 
\midrule 
Total & 0.0321 & 0.0205 & 0.051\\ 
\bottomrule\end{tabular} 
\end{center} 
\end{table}

\section{Results and conclusion}
\label{sec:Results}

Using a fit to the $m(D^{*-}D_s^+)$ distribution to determine the properties of partially reconstructed \mbox{$B^0 \to D^{*-} (D_s^{*+} \to D_s^+ \gamma)$} decays, the longitudinal polarisation fraction is measured to be
\begin{equation*}
\fL = \fLsyst,
\end{equation*}
where the first uncertainty is statistical and the second is systematic. The corresponding magnitude of the longitudinal helicity amplitude, given by $|H_0| = \sqrt{f_{\rm L}}$, is
\begin{equation*}
|H_0| = \Hzamp.
\end{equation*}
This information is used to measure the remaining helicity observables in an angular fit to fully reconstructed \mbox{$B^0 \to D^{*-} (D_s^{*+} \to D_s^+ \gamma)$} decays, obtaining
\begin{align*}
|H_-| &= \phantom{-}\Hmamp, \\
|H_+| &= \phantom{-}\Hpamp, \\
\phi_+ &= \Hpphi, \\
\phi_- &= \phantom{-}\Hmphi,
\end{align*}
where the quoted value and uncertainties for $|H_+|$ are fully determined by the normalisation of the three helicity amplitudes to unity. The measurement of \fL is consistent with and more precise than the current world average, $\fL = 0.52 \pm 0.05$~\cite{Aubert:2003jj,PDG2020}. The transverse amplitude magnitudes and phases are measured for the first time, where both phases are consistent with zero but the magnitudes differ from each other at the level of nine standard deviations. It is noted that $|H_0| > |H_+| > |H_-|$, which is expected from quark-helicity
conservation in $B$ decays involving a $b \to c$ quark transition. In such decays, the $V-A$ nature of the weak interaction causes the
longitudinal component to dominate. The inequality is stronger for decays involving light vector mesons~\cite{Suzuki:2001za}, but also appears to be satisfied in $B^0 \to D^{*-} D_s^{*+}$ decays where two vector charm mesons are produced. This helicity hierarchy is not observed in decays dominated by penguin amplitudes such as $B^0 \to \phi K^{*0}$, where the longitudinal
and transverse components are found to have roughly equal amplitudes~\cite{LHCb-PAPER-2014-005,delAmoSanchez:2010mz,Abe:2004mq,LHCb-PAPER-2011-012}.

The branching fraction ratio of \mbox{$B^0 \to D^{*-} (D_s^{*+} \to D_s^+ \gamma)$} decays relative to \mbox{$B^0 \to D^{*-} D_s^+$} decays is measured to be
\begin{equation*}
\mathcal{R} = \BFratio, 
\end{equation*}
where the first uncertainty is statistical and the second is systematic. This result is in agreement with, but considerably more precise than, the current world-average value $\mathcal{R} = 2.07 \pm 0.33$~\cite{PDG2020}. The branching fraction ratio of the Cabibbo-suppressed \mbox{$B_s^0 \to D^{*-} D_s^+$} decay relative to the \mbox{$B^0 \to D^{*-} D_s^+$} decay is measured to be
\begin{equation*}
r(\Bs) = \BsBFratio,  
\end{equation*}
where the first uncertainty is statistical, the second is systematic, and the third accounts for the use of an external value of $f_s/f_d$~\cite{LHCb-PAPER-2020-046}. This measurement constitutes the first observation of the Cabibbo-suppressed $\Bs \to D^{*-} D_s^+$ decay with a significance of seven standard deviations.

In conclusion, an angular analysis of \BdDstDsst  with $D_s^{*+} \to D_s^+ \gamma$ decays is performed using 6\invfb of data collected with the LHCb experiment at $\sqrt{s} = 13$\tev in order to measure a complete set of helicity amplitude observables. Partially reconstructed candidates are used in a fit to the $m(D^{*-}D_s^+)$ distribution to measure the longitudinal polarisation fraction $\fL = |H_0|^2$. This knowledge is then used in a subsequent angular fit to fully reconstructed data in order to measure the remaining helicity observables. The measurement of \fL is consistent with and more precise than the current world-average value, while the magnitudes and phases of the transverse helicity amplitudes are measured for the first time. The pattern of helicity amplitude magnitudes is found to align with expectations from quark-helicity conservation for tree-level $B$ decays involving a $b \to c$ transition. The \BdDstDsst decay is a large background in \BdDstTauNu analyses, particularly when the $\tau^+$ decays hadronically. Analyses aiming to measure angular observables in \BdDstTauNu decays must control the angular distributions of prominent hadronic backgrounds such as \BdDstDsst, and the results presented herein will help to significantly reduce background model uncertainties in future measurements.

\section*{Acknowledgements}
%
%
\noindent We express our gratitude to our colleagues in the CERN
accelerator departments for the excellent performance of the LHC. We
thank the technical and administrative staff at the LHCb
institutes.
We acknowledge support from CERN and from the national agencies:
CAPES, CNPq, FAPERJ and FINEP (Brazil); 
MOST and NSFC (China); 
CNRS/IN2P3 (France); 
BMBF, DFG and MPG (Germany); 
INFN (Italy); 
NWO (Netherlands); 
MNiSW and NCN (Poland); 
MEN/IFA (Romania); 
MSHE (Russia); 
MICINN (Spain); 
SNSF and SER (Switzerland); 
NASU (Ukraine); 
STFC (United Kingdom); 
DOE NP and NSF (USA).
We acknowledge the computing resources that are provided by CERN, IN2P3
(France), KIT and DESY (Germany), INFN (Italy), SURF (Netherlands),
PIC (Spain), GridPP (United Kingdom), RRCKI and Yandex
LLC (Russia), CSCS (Switzerland), IFIN-HH (Romania), CBPF (Brazil),
PL-GRID (Poland) and NERSC (USA).
We are indebted to the communities behind the multiple open-source
software packages on which we depend.
Individual groups or members have received support from
ARC and ARDC (Australia);
AvH Foundation (Germany);
EPLANET, Marie Sk\l{}odowska-Curie Actions and ERC (European Union);
A*MIDEX, ANR, IPhU and Labex P2IO, and R\'{e}gion Auvergne-Rh\^{o}ne-Alpes (France);
Key Research Program of Frontier Sciences of CAS, CAS PIFI, CAS CCEPP, 
Fundamental Research Funds for the Central Universities, 
and Sci. \& Tech. Program of Guangzhou (China);
RFBR, RSF and Yandex LLC (Russia);
GVA, XuntaGal and GENCAT (Spain);
the Leverhulme Trust, the Royal Society
 and UKRI (United Kingdom).

\clearpage

\section*{Appendices}

\appendix

\section{Relationship between $m(D^{*-}D_s^+)$ and $\cos\theta_X$}
\label{app:m_DstDs_costheta_X}

\begin{figure}[!h]
  \begin{center}
   \includegraphics[width=0.65\linewidth]{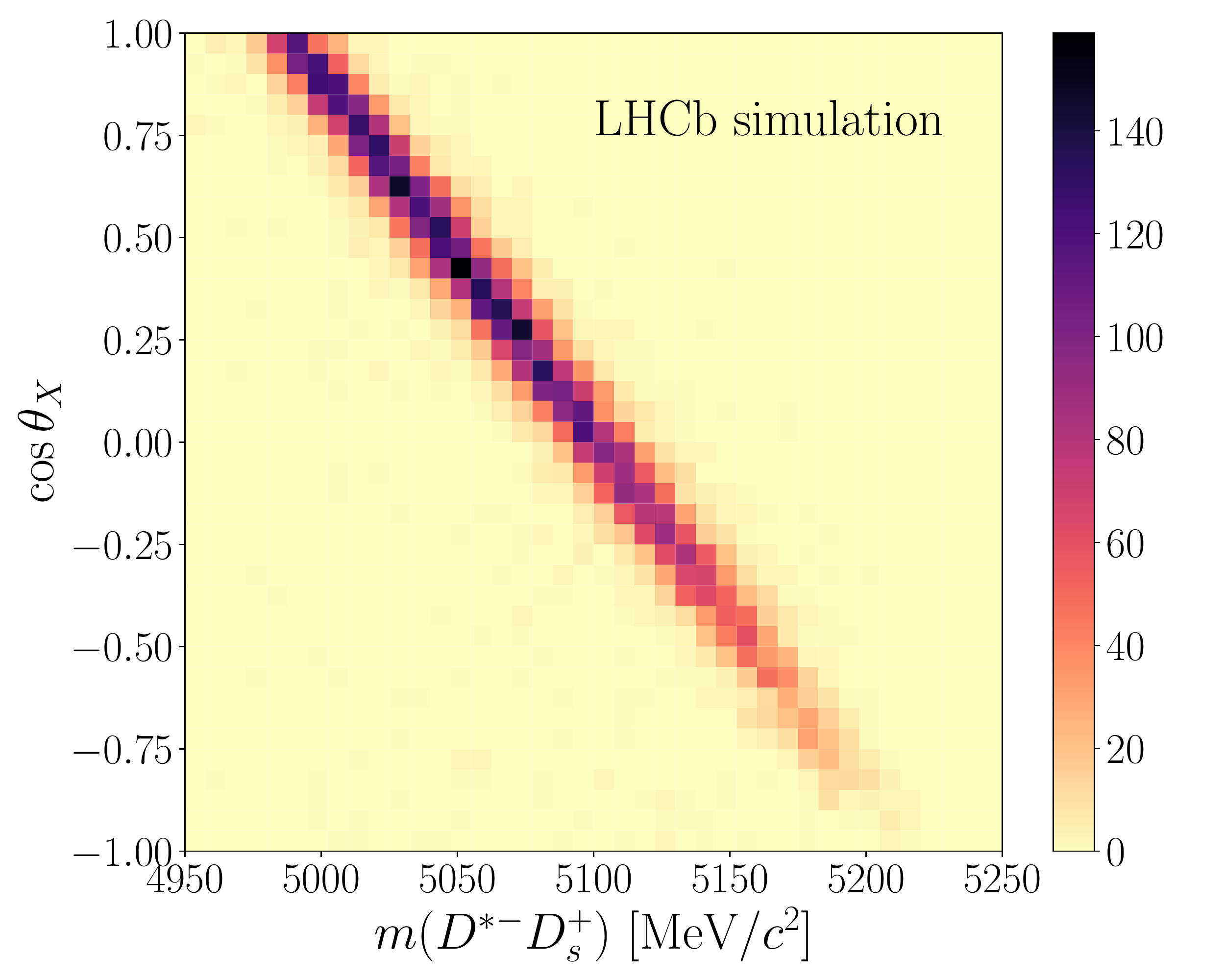}
   \end{center}
   \caption{Relationship between $m(D^{*-}D_s^+)$ and $\cos\theta_X$ in a sample of fully reconstructed \mbox{$B^0 \to D^{*-} (D_s^{*+} \to D_s^+ \gamma)$} simulated decays. The colour scale indicates the number of candidates in each bin.
  \label{fig:m_DstDs_vs_costheta_X}}
\end{figure}

\begin{figure}[!h]
  \begin{center}
   \includegraphics[width=0.5\linewidth]{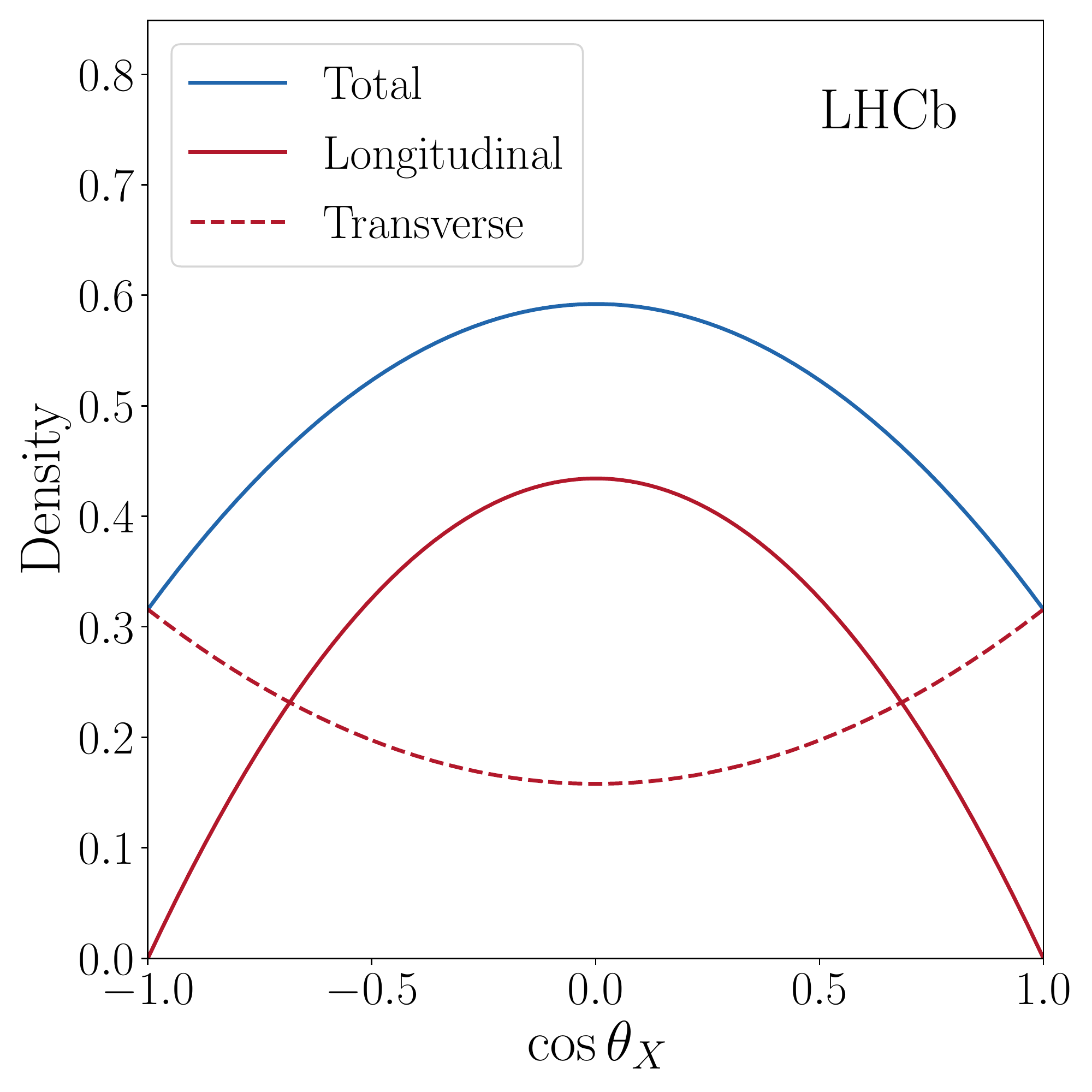}
   \end{center}
   \caption{Transverse and longitudinal contributions to the one-dimensional decay rate shown as a function of $\cos\theta_X$.
  \label{fig:costheta_X_1D_plots}}
\end{figure}

\begin{figure}[!h]
  \begin{center}
   \includegraphics[width=0.49\linewidth]{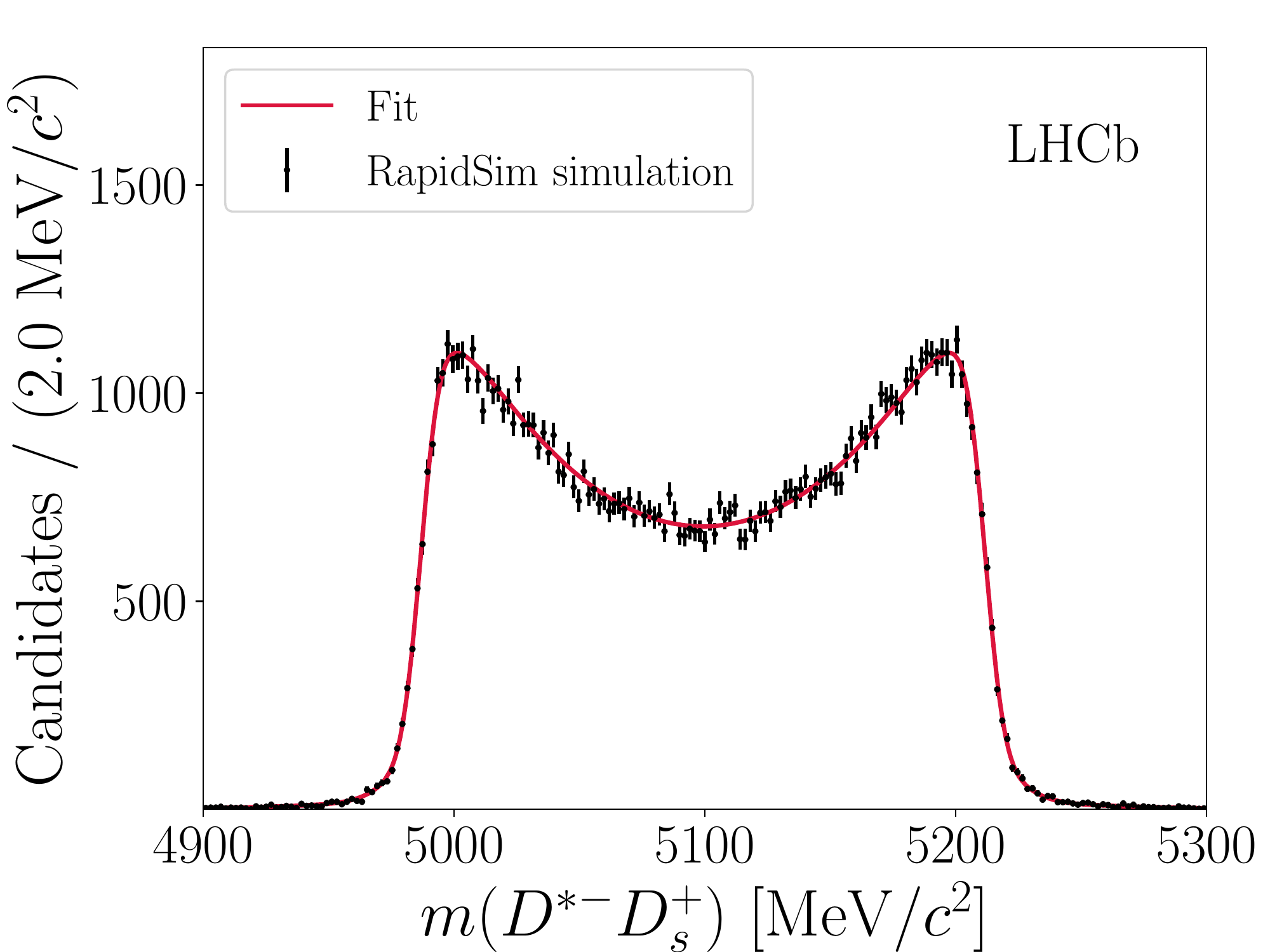} 
   \includegraphics[width=0.49\linewidth]{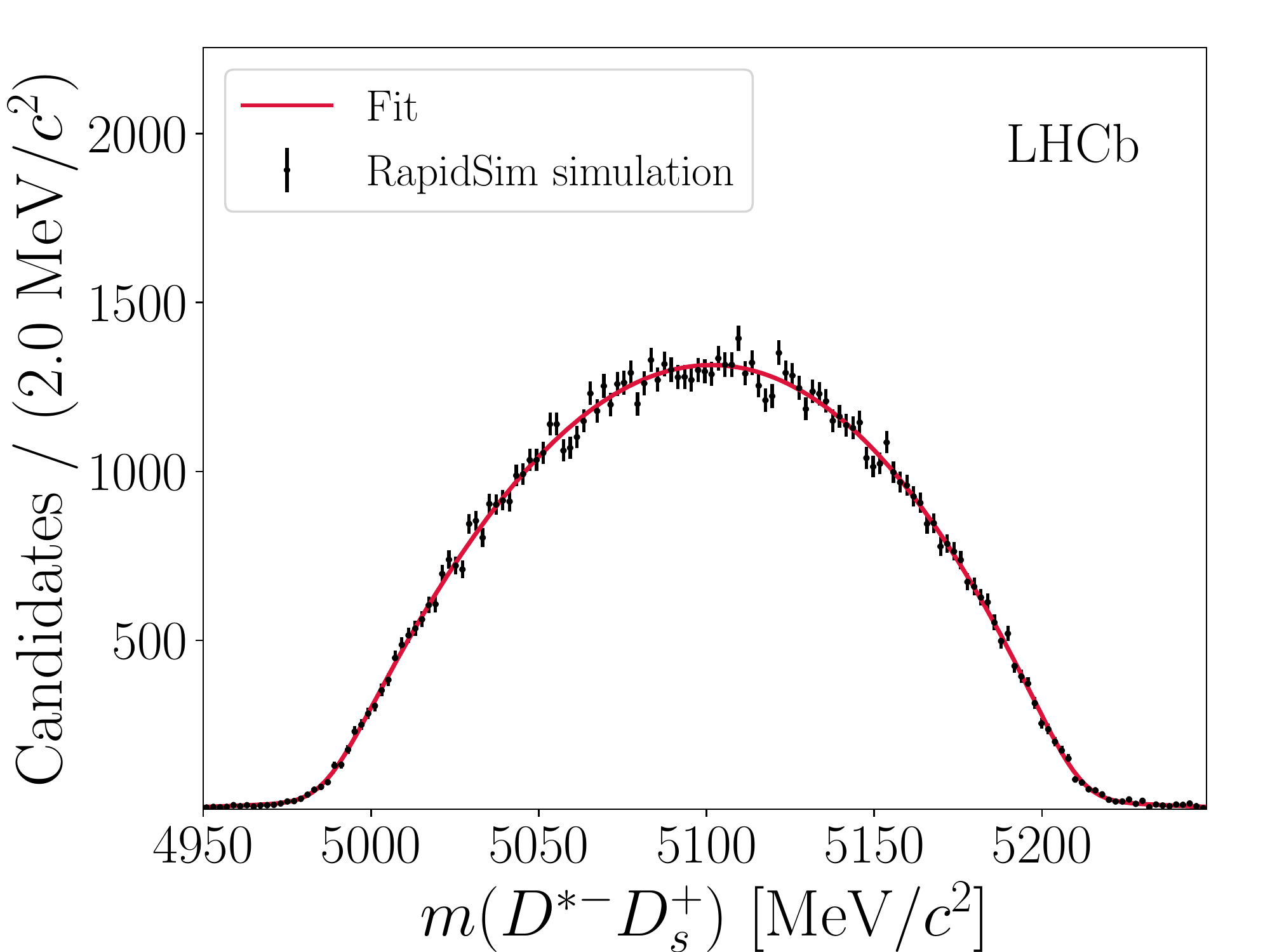}
   \end{center}
   \caption{Invariant-mass distributions of (left) pure transverse and (right) longitudinal \mbox{$B^0 \to D^{*-} (D_s^{*+} \to D_s^+ \gamma)$} simulated decays. Fits to the distributions are overlaid, from which shape parameters for use in the $m(D^{*-}D_s^+)$ data fit are derived.
  \label{fig:RapidSim_fits}}
\end{figure}

In Fig.~\ref{fig:m_DstDs_vs_costheta_X}, the relationship between $m(D^{*-}D_s^+)$ and $\cos\theta_X$ is shown for fully reconstructed $B^0 \to D^{*-} (D_s^{*+} \to D_s^+ \gamma)$ simulated decays. A strong negative correlation is evident, due to a common dependence on the kinematics of the photon produced in the $D_s^{*+}$ decay. The one-dimensional decay rate as a function of $\cos\theta_X$ is given by Eq.~(\ref{eq:costheta_X_decay_rate}), where separate transverse and longitudinal components contribute; these components are illustrated in Fig.~\ref{fig:costheta_X_1D_plots}. Due to the co-dependence of  $m(D^{*-}D_s^+)$ and $\cos\theta_X$, the different angular forms for transverse and longitudinal signal give rise to different $m(D^{*-}D_s^+)$ distributions. This is illustrated in Fig.~\ref{fig:RapidSim_fits}, where \textsc{RapidSim} samples of transverse and longitudinal signal are shown. The fits used to derive shape parameters for the $m(D^{*-}D_s^+)$ fit are overlaid.

\FloatBarrier



\FloatBarrier


\addcontentsline{toc}{section}{References}
\bibliographystyle{LHCb}
\bibliography{main,standard,LHCb-PAPER,LHCb-CONF,LHCb-DP,LHCb-TDR}

\newpage
\centerline
{\large\bf LHCb collaboration}
\begin
{flushleft}
\small
R.~Aaij$^{32}$,
C.~Abell{\'a}n~Beteta$^{50}$,
T.~Ackernley$^{60}$,
B.~Adeva$^{46}$,
M.~Adinolfi$^{54}$,
H.~Afsharnia$^{9}$,
C.A.~Aidala$^{86}$,
S.~Aiola$^{25}$,
Z.~Ajaltouni$^{9}$,
S.~Akar$^{65}$
% ...

{\footnotesize \it

% ...
$^{9}$Universit{\'e} Clermont Auvergne, CNRS/IN2P3, LPC, Clermont-Ferrand, France\\
% ...
$^{25}$INFN Sezione di Milano, Milano, Italy\\
% ...
$^{32}$Nikhef National Institute for Subatomic Physics, Amsterdam, Netherlands\\
% ...
$^{46}$Instituto Galego de F{\'\i}sica de Altas Enerx{\'\i}as (IGFAE), Universidade de Santiago de Compostela, Santiago de Compostela, Spain\\
% ...
$^{50}$Physik-Institut, Universit{\"a}t Z{\"u}rich, Z{\"u}rich, Switzerland\\
% ...
$^{54}$H.H. Wills Physics Laboratory, University of Bristol, Bristol, United Kingdom\\
% ...
$^{60}$Oliver Lodge Laboratory, University of Liverpool, Liverpool, United Kingdom\\
% ...
$^{65}$University of Cincinnati, Cincinnati, OH, United States\\
% ...
$^{68}$Syracuse University, Syracuse, NY, United States\\
% ...
$^{86}$University of Michigan, Ann Arbor, United States, associated to $^{68}$\\
\bigskip
% ...
\medskip
}
\end{flushleft}

\end{document}